\documentclass[aps,nofootinbib,superscriptaddress,10pt,floatfix]{revtex4}
\normalsize
\usepackage{mathrsfs}
\usepackage[scr=boondox]{mathalpha}  
\usepackage{graphicx}
\usepackage{amsmath,amssymb}
\usepackage{braket}
\usepackage{subfigure}
\usepackage{subfloat}
\usepackage{float}
\usepackage{mciteplus}
\usepackage[usenames]{color}
\usepackage[normalem]{ulem}
\usepackage{comment}
\usepackage{ifthen} 
\usepackage{fp,calc}
\usepackage{lastpage}
\usepackage{setspace}
\usepackage[pdftex]{epsfig}
\DeclareGraphicsExtensions{.jpg,.mps,.pdf,.png}
\usepackage{cellspace}
\usepackage{booktabs}
\usepackage{latexsym}
\usepackage{graphics}
\usepackage{fancyhdr}
\usepackage{adjustbox}
\usepackage[english]{babel}
\usepackage[svgnames]{xcolor}
\usepackage{enumitem}
\usepackage{tabularx}

\newcommand{\bn}{\mathbf{n}}

\newcommand{\dd}{{\rm d}}
\newcommand{\E}{{\rm e}}
\newcommand{\M}{\mathcal{M}}

\def\beq{\begin{equation}}
\def\eeq{\end{equation}}

\newcommand{\rbr}[1]{\left(#1\right)}
\newcommand{\sbr}[1]{\left[#1\right]}
\newcommand{\angbr}[1]{\left\langle#1\right\rangle}

\newcommand{\cond}{_{h\vert\rho_g}}

\newcommand{\st}{{\mathrm{st.}}}
\newcommand{\clust}{{\mathrm{clust.}}}

\newcommand{\be}{\begin{equation}}
\newcommand{\ee}{\end{equation}}
\newcommand{\cat}{\cite{BarausseCatalogue, Klein:2015hvg}}

\usepackage{float}
\usepackage{graphicx}

\newboolean{articletitles}
\setboolean{articletitles}{true} 
\newboolean{allauthors}
\setboolean{allauthors}{false} 

\mciteErrorOnUnknownfalse
\usepackage[colorlinks=true, allcolors=blue]{hyperref}

\definecolor{Mgreen}{rgb}{0.1, 0.69,0.16}

\begin{document}

\title{The impact of large-scale galaxy clustering on the variance of the Hellings-Downs correlation: numerical results}

\author{Nastassia Grimm}
\email[]{nastassia.grimm@unige.ch}
\affiliation{D\'epartement de Physique Th\'eorique and Center for Astroparticle Physics, Universit\'e de Gen\`eve, Quai E. Ansermet 24, CH-1211 Geneve 4, Switzerland}

\author{Martin Pijnenburg}
\email[]{martin.pijnenburg@unige.ch}
\affiliation{D\'epartement de Physique Th\'eorique and Center for Astroparticle Physics, Universit\'e de Gen\`eve, Quai E. Ansermet 24, CH-1211 Geneve 4, Switzerland}

\author{Giulia Cusin}
\email[]{giulia.cusin@iap.fr}
\affiliation{Institut d'Astrophysique de Paris, UMR-7095 du CNRS, Paris, France}
\affiliation{D\'epartement de Physique Th\'eorique and Center for Astroparticle Physics, Universit\'e de Gen\`eve, Quai E. Ansermet 24, CH-1211 Geneve 4, Switzerland}

\author{Camille Bonvin}
\email[]{camille.bonvin@unige.ch}
\affiliation{D\'epartement de Physique Th\'eorique and Center for Astroparticle Physics, Universit\'e de Gen\`eve, Quai E. Ansermet 24, CH-1211 Geneve 4, Switzerland}

\date{\today}

\begin{abstract}

Pulsar timing array experiments have recently found evidence for a stochastic gravitational wave (GW) background, which induces correlations among pulsar timing residuals  described by the Hellings and Downs (HD) curve. Standard calculations of the HD correlation and its variance assume an isotropic background. However, for a background of astrophysical origin, we expect a higher GW spectral density in directions with higher galaxy number densities. In a companion paper, we have developed a theoretical formalism to account for the anisotropies arising from large-scale galaxy clustering, leading to a new contribution to the variance of the HD correlation. In this subsequent work, we provide numerical results for this novel effect. We consider a GW background resulting from mergers of supermassive black hole binaries, and relate the merger number density to the overdensity of galaxies.  We find that anisotropies due to large-scale galaxy clustering lead to a standard deviation of the HD correlation at most at percent level, remaining well below the standard contributions to the HD variance. 
Hence, this kind of anisotropies in the GW source distribution does not represent a substantial contamination to the correlations of timing residuals in present and future PTA surveys. Suitable statistical methods to extract the galaxy clustering signal from PTA data will be investigated in the future.  

\end{abstract}

\maketitle

\tableofcontents

\newpage

\section{Introduction}

Pulsar timing arrays (PTAs) were used to deliver the first evidence of a stochastic gravitational wave (GW) background in the nHz band~\cite{NANOGrav:2023gor, Reardon:2023gzh, EPTA:2023sfo, Xu:2023wog}.  The correlation of pulsar timing residuals due to the passing of GWs, as a function of the pulsar pair separation on the sky, is usually assumed to be described by the Hellings and Downs (HD) curve~\cite{Hellings:1983fr}.\footnote{We note that we use the notion \textit{HD curve} to refer to the theoretical prediction by Hellings and Downs~\cite{Hellings:1983fr}, while \textit{HD correlation} corresponds to the observed correlation that is in fact measured in our Universe and might differ from the idealized HD curve.}  However, it has been recently shown that the stochasticity of GW sources, and the fact that the number of pulsar pairs at a given separation is finite, cause a departure from this idealized curve~\cite{Allen:2022bjz, Allen:2022dzg, Allen:2022ksj, Romano:2023zhb, Allen:2024rqk, Bernardo:2022xzl}. In these studies, however, the GW sources are assumed to be isotropically distributed. In Ref.~\cite{Grimm:2024lfj} we presented a novel framework to account for the impact of the cosmological large-scale structure on the HD correlation.\footnote{Prior to our work in Ref.~\cite{Grimm:2024lfj}, the possibility of anisotropies in the distribution of GW sources in the PTA band has been pointed out in various studies, however either not in the context of the HD correlation \cite{Allen:1996gp,Cusin:2017fwz, Cusin:2017mjm, Cusin:2018rsq, Cusin:2018avf, Cusin:2019jpv, Cusin:2019jhg, Pitrou:2019rjz, Jenkins:2019nks, Alonso:2020mva, renzini2022}, or without targeting the impact on its variance \cite{Mingarelli:2013dsa, Taylor:2013esa, Gair:2014rwa, Ali-Haimoud:2016mbv}. The detectability of kinematic anisotropies in the SGWB with different PTA experiments has been recently studied in Ref.~\cite{Cruz:2024svc}.} Indeed, since galaxies are not isotropically distributed on the sky but cluster to form the large-scale structure, we expect an intrinsic isotropy to be present as well for astrophysical GW sources.

 The timing residual measured for a given pulsar is linear in the total GW strain in a given direction, due to a superposition of GWs, emitting incoherently by a large number of sources. Such a signal is stochastic for two distinct reasons: first because we cannot predict the properties of the emitted waves (in particular the phases), and second because we cannot predict the distribution of sources. Usually, the second source of stochasticity is ignored: sources are assumed to be isotropically distributed, and there is therefore only one possible realization of the distribution of sources. The HD curve is then defined as the product of the timing residuals of two pulsars, averaged over all realizations of the GW emission. 

As we have shown in Ref.~\cite{Grimm:2024lfj}, to account for anisotropies in the distribution of sources due to large-scale galaxy clustering, one needs to add an additional average over all possible realizations of the distribution of sources. We  considered an ensemble of GW sources in one realization of the universe (i.e.\ a fixed distribution of sources), and then analyzed how the HD correlation varies from realization to realization. The mean of the HD correlation itself is not affected by clustering, which means that anisotropies in the distribution of sources do not generate a systematic bias. However, the large-scale structure of the Universe induces a variance in the measurement of the HD correlation, quantifying to which amount the measurements in our specific realization of the Universe can differ from the mean.\footnote{In this work and in the companion paper \cite{Grimm:2024lfj} we consider only the effect of clustering (i.e.\ fluctuations in the distribution of sources) on the variance of the HD correlation, neglecting relativistic effects (due to gravitational potentials and source velocities) that are expected to be subdominant on all scales, see e.g.~the calculation in Ref.~\cite{Cusin:2017fwz, Pitrou:2019rjz} in the context of ground-based detectors. } 

In this work, we numerically evaluate the size of this variance. To do so, we assume that the observed background is given by the GW emission of supermassive black hole (SMBH) binaries in the inspiraling phase. 
For our case study we choose to work with the catalogs of Refs.~\cat, and we show results for the three population models considered therein: the {\tt popIII} model based on light growth seeds for SMBHs; and the other two heavy-seed {\tt Q3} SMBH  models, with ({\tt Q3-d}) and without ({\tt Q3-nd}) delays between galaxy merger and SMBH binary merger.  We assume that the overdensity of black hole mergers is a biased tracer of the underlying galaxy overdensity, which itself is a biased tracer of the matter distribution. We find that anisotropies due to large-scale galaxy clustering lead to a departure from the HD curve at a sub-percent level, ranging from 0.54\% to 2.85\% depending on pulsar pair separation, i.e.~well below the precision of current PTA surveys~\cite{NANOGrav:2023gor, Reardon:2023gzh, EPTA:2023sfo, Xu:2023wog}. Moreover, it is well below the standard variance calculated in Ref.~\cite{Allen:2022dzg} under the assumption of an isotropic source distribution. This indicates that the clustering of sources can safely be neglected in current and future analyses. 

In addition to these results, which refer to a single pulsar pair, we also compute the variance related to an (infinite) continuous distribution of pulsar pairs, which we refer to as the \textit{pulsar-averaged} case.\footnote{In a previous version of this manuscript, we have referred to the pulsar-averaged as the irreducible case. However, using the term \textit{irreducible} in this context is misleading. As recently shown in Ref.~\cite{Pitrou:2024scp} (and further supported by Ref.~\cite{Allen:2024uqs} for a real-space result), cosmic variance can be mitigated through appropriate frequency weighting. We likewise expect that the clustering-induced variance can also be reduced using a similar weighting approach.} Averaging over pairs of pulsars reduces the clustering variance by an order of magnitude, since anisotropies in the distribution of sources affect pulsar pairs at different sky locations differently and are therefore washed out by the average over an infinite number of pulsars. The pulsar-averaged and single pair cases provide, respectively, a lower and upper bound on the clustering variance in a real observation with a finite number of pulsar pairs. 

We compare our results with the approximate computations of the clustering variance derived in Ref.~\cite{Allen:2024mtn} for the pulsar-averaged case. We find that the pulsar-averaged clustering variance is, with a relative amplitude of $0.10-0.14\%$, an order of magnitude smaller than the upper bound of 1\% claimed in Ref.~\cite{Allen:2024mtn}.  
This difference is due to the different modeling of the clustering correlation of GW sources. Ref.~\cite{Allen:2024mtn} assumes a constant value for the angular power spectrum of galaxy correlations (independent on angular separation) and does not take any redshift dependence into account. In contrast, we are using the clustering correlation predicted in a $\Lambda$CDM cosmological model, which properly accounts for the non-trivial dependence of clustering correlations on source angular separation and on redshift. Our formalism provides therefore a clear link between our cosmological and astrophysical modeling and the variance of the HD correlation.

The remaining part of this paper is structured as follows. In Section~\ref{sec:Basics}, we describe basic concepts concerning PTAs and galaxy clustering in the context of this work. We also give an overview over different types of variance contributions to the HD correlation introduced in this and previous works~\cite{Grimm:2024lfj, Allen:2022dzg}, and summarize the analytical results of Ref.~\cite{Grimm:2024lfj} for the clustering variance. Then, in Section~\ref{sec:ast_modeling}, we present the modeling and catalog reconstruction of the quantity $b_{\rm GW}$, which relates galaxy density to GW spectral density and modulates the amplitude of the clustering variance. The fact that SMBH mergers are biased tracers of the underlying galaxy and matter fields are discussed as well. Section~\ref{sec:results} presents the numerical results for the clustering variance obtained in this work. We also describe our numerical methods and provide a comparison to Ref.~\cite{Allen:2024mtn} in this section. Finally, we conclude in Section~\ref{sec:conclusion}. Moreover, in Appendix~\ref{App:Shot_pulsar_variance}, we present the analytical results for the pulsar-averaged clustering variance in the case of infinitely many pulsar pairs.

\section{Basic concepts related to the Hellings and Downs correlation} \label{sec:Basics}

The theoretical modeling of the HD correlation and its variance, accounting for anisotropies in the distribution of sources, has been developed in Ref.~\cite{Grimm:2024lfj}. Here, we summarize the key concepts and notations.

\subsection{Description of the signal}

The passing of GWs induces redshifts in pulsar observations, i.e.~small shifts in the arrival period between two pulses.\footnote{To avoid confusion, we denote with a capital $Z$ the redshift of pulsars induced by the passing of GWs, and with $z$ the cosmological redshift (of the sources) due to the Universe's expansion. When not explicitly specified, the term ``redshift'' refers to the latter.} We write the GW propagating in direction $\bn$ (due to the sum of waves from various sources propagating in that direction) as 
\be
h_{ij}(t, \bn)\equiv  \int {\rm d}f \sum_{A=+, \times} e_{ij}^A(\bn) h_A(f, \bn) \E^{2\pi i f t}\,, \label{Eq:hij}
\ee
where $e_{ij}^A(\bn)$ is the wave polarization tensor. The pulsar redshift $Z_a$, for a pulsar located in direction $\bn_a$, is given by (see e.g.\ Ref.~\cite{Allen:2022dzg}),
\be
Z_a(t)\equiv \sum_A \int {\rm d}f \int \dd \bn\, F_a^A(\bn)h_A(f, \bn) \E^{2\pi i f t}\left(1-\E^{-2\pi i 
 f \tau_a (1+\bn\cdot{\bf n}_a)}\right)\,, \label{Eq:Redshift}
\ee
where $\tau_a$ is the light travel time between emission and observation. Furthermore, the antenna pattern functions $F_a^A(\bn)$ are defined as  
\be\label{antenna}
F_a^A(\bn)=\frac{n_a^i n_a^j e_{ij}^A(\bn)}{2(1+\bn\cdot \bn_a)}\,.
\ee
The pulsar redshift is directly proportional to the amplitude (or strain) of the GWs, $h_A(f,\bn)$, where $A$ denotes the polarization of the wave, $+$ and $\times$.

The aim of PTA observations is to detect correlated shifts in the pulses of two distinct pulsars, due to the passing of GWs. We therefore introduce the quantity $\rho_{ab}$, defined as the product of $Z_a$ and $Z_b$ for two pulsars located at $\bn_a$ and $\bn_b$, averaged over the observation time $T$,
\be 
\rho_{ab}\equiv\overline{Z_aZ_b}\,. \label{Eq:rho_ab}
\ee
The HD curve describes the theoretical expectation of this quantity, and it is obtained by taking the ensemble average of Eq.~\eqref{Eq:rho_ab} accounting for both the stochasticity of GWs and the fact that sources are distributed in a stochastic way. More details on these different layers of stochasticity can be found in Ref.~\cite{Grimm:2024lfj}. Therein, we have used distinct notations to highlight the difference between full ensemble averages and ensemble averages assuming a given density distribution. In this work, as we focus on the numerical predictions, we omit this distinction in our notation and refer to our previous work for the formal details.  

\subsection{The Hellings and Downs correlation in the clustered Universe}

PTA surveys measure the angular correlations of pulsar redshifts, with theoretical expectation value given by $\langle\rho_{ab}\rangle$. Here, the average is in fact a double average:  a first average over GW realizations and a second average over stochastic initial conditions of the matter overdensity (of which the distribution of GW sources is a biased tracer). From Eqs.~\eqref{Eq:Redshift} and~\eqref{Eq:rho_ab}, we see that the quantity $\langle\rho_{ab}\rangle$ contains ensemble averages over the product of two $h_A(f,\bn)$. For a given realization of the density field, this ensemble average over GW realizations is given by 
\begin{equation}
\left\langle h_A^\ast(f,\mathbf{n})h_B(f',\mathbf{n'})\right\rangle =\frac 12 \delta_{AB}\,\delta(f-f') S_h(f,\bn)\frac{\delta^2(\mathbf{n},\mathbf{n'})}{4\pi}\,, \label{eq:hh}
\end{equation} 
where the delta functions ensure that only GWs with the same polarization state, at the same frequency and from the same direction are correlated.\footnote{This is another way of stating that, when we sum over a large number of products of incoherent signals, non-vanishing
contributions arise only from products of signals with equal phases, see e.g.\ Ref.~\cite{Cusin:2017fwz}.} The spectral density $S_h(f,\bn)$ is usually assumed to be isotropic, $S_h(f,\bn) = \bar S_h(f)$. Then, the HD correlation is determined by
\be
\langle\rho_{ab}\rangle=h^2\mu_{\rm HD}(\gamma)\,,\qquad h^2=\int\mathrm df\,\bar S_h(f)\,, \label{Eq:HD_strain}
\ee
where $h^2$ is the squared strain, $\gamma$ denotes the angle between the pulsar directions $\bn_a$ and $\bn_b$, and the HD curve $\mu_{\rm HD}(\gamma)$ is given by
\begin{equation}
 \mu_{\rm HD}(\gamma)=\int\frac{\mathrm d\bn}{4\pi}\chi_{ab}(\bn)=\frac 14+\frac 1{12}\cos(\gamma)+\frac 12 (1-\cos(\gamma))\log\left(\frac{1-\cos(\gamma)}{2}\right)\,. \label{Eq:HDcurve}
\end{equation}
Indeed, due to the statistical isotropy of $h_A$, the HD curve depends on $\bn_a$ and $\bn_b$ only via their angular separation $\gamma$. More specifically, it is determined via the sky-average of the geometrical objects $\chi_{ab}(\bn)$,
\beq
\chi_{ab}(\bn)=\sum_{A=+, \times} F_a^A(\bn)F_b^A(\bn)\,.
\eeq

In Ref.~\cite{Grimm:2024lfj} and in this work, we account for the fact that galaxies follow the cosmological large-scale structure. This leads to higher galaxy densities and therefore, in case of an astrophysical origin of the stochastic GW background, to higher spectral densities in certain directions. Splitting $S_h(f,\bn)$ into an isotropic and anisotropic part, we can relate it to the fractional galaxy overdensity $\delta_g(\bn,z)$ as
\beq
S_h(f,\mathbf{n})=\bar S_h(f)+\delta S_h(f,\mathbf{n})=\bar S_h(f)\sbr{1+\int\mathrm d z\,b_{\rm GW}(z)\delta_g(\mathbf{n},z)}\,, \label{eq:sh_clust}
\eeq
where $b_{\rm GW}(z)$ is some redshift dependent factor\footnote{We note that in Ref.~\cite{Grimm:2024lfj}, to be as general as possible, we had introduced $b_{\rm GW}(f,z)$ as a redshift and frequency dependent factor. In this work, however, we will see that $b_{\rm GW}(z)$ appears as a frequency independent factor in our catalog-based astrophysical modeling.} that relates GW anisotropies to galaxy anisotropies. It depends on the details of the astrophysical model in consideration and we will compute it below for the case where the SGWB is due to SMBH mergers. We note that $b_{\rm GW}(z)$ does not depend on the direction $\bn$: since we assume that the distribution of GW sources follows the distribution of galaxies, the anisotropic behavior of $S_h(f,\bn)$ is determined by $\delta_g(\bn,z)$ alone.

To obtain the expectation value of the HD correlation while taking galaxy clustering into account, we need to take the ensemble average of Eq.~\eqref{eq:hh} over all realizations of the galaxy density field $\delta_g$. In Ref.~\cite{Grimm:2024lfj}, we showed that since $\rho_{ab}$ is linear is $\delta_g$ and since this quantity has zero mean, the anisotropic part $\delta S_h(f,\bn)$ vanishes when averaged over all realizations. The mean of the HD correlation is therefore unaffected by anisotropies in the distribution of sources. This is, however, not true for the variance of the HD correlation, which obtains an additional contribution due to galaxy clustering. In the following, we will summarize different types of variances, before discussing analytical and numerical results in the subsequent sections.

\subsection{Different types of variance}

The aim of our work is to provide a complete computation of the variance of the HD correlation, i.e.~to assess by how much a measurement of angular correlations of pulsar redshifts in a single realization of the universe can differ from the expectation value $\langle\rho_{ab}\rangle$, 
taking all sources of variance in the presence of anisotropies into account. We distinguish different types of variances, due to the different sources of stochasticity in the signal. 

The standard variance of the HD correlation computed in Refs.~\cite{Allen:2022bjz, Allen:2022dzg, Allen:2022ksj, Romano:2023zhb, Allen:2024rqk} is evaluated assuming a fixed, isotropic realization of the distribution of sources, i.e.~neglecting the stochastic nature of the source distribution. It is related to the 4-point function of the strain averaged over all GW realizations. We refer to this variance as \emph{standard variance}. It describes how the signal can differ from the HD curve due to the fact that we observe only one realization of the GW strain.  In contrast, in Ref.~\cite{Grimm:2024lfj}, we also accounted for the stochasticity in the source distribution. We refer to the resulting variance as \emph{clustering variance}. It describes how the signal can differ from the HD curve due to the fact that we observe only one realization of the density field. 

In Eq.~\eqref{Eq:rho_ab}, the signal is computed for a single pair of pulsars $a$ and $b$. In practice however, one can average over all pairs of pulsars in the array that share the same angular separation $\gamma$. This reduces the variance, since anisotropies in the GW signal affect different pulsar pairs (in different directions) differently. Averaging over many pairs therefore erases part of the anisotropies. The reduction of the variance depends on the number of pulsar pairs per angular separation $\gamma$.\footnote{Investigations of the standard variance for various finite sets of pulsar pairs have been performed in Ref.~\cite{Allen:2022ksj}. In particular, Fig.~9 therein shows that, for 88 pulsars corresponding to the specifications of the International Pulsar Timing Array~\cite{NANOGrav:2023gor, Reardon:2023gzh, EPTA:2023sfo, Xu:2023wog}, the variance is already considerably reduced compared to the case of only one pair per angular separation. Future data sets, containing an even larger number of pulsars, will approach the pulsar-averaged limit (with an infinite set of pulsars) even more. We stress again that this limit can be further reduced with a proper frequency weighting, as recently shown in Ref.~\cite{Pitrou:2024scp}. } The limiting case, corresponding to a continuous (infinite) distribution of pulsars, provides a lower bound to the variance. We call these variances \emph{pulsar-averaged standard variance} and \emph{pulsar-averaged clustering variance}. A summary of the different contributions to the variance of the HD correlation is presented in Table~\ref{TableBig}. In the main part of this work, we show the computation of the clustering variance in the case of a single pulsar pair per angular separation $\gamma$ (which provides an upper bound). Results for the pulsar-averaged variance, obtained by averaging over a continuous, infinite number of pulsars, will be derived and presented in Appendix \ref{App:Shot_pulsar_variance}. 

Finally, we note that in principle, an additional shot noise contribution arises from the fact that the signal is due to a discrete number of GW sources (see, e.g., Ref.~\cite{Jenkins:2019nks}). This contribution accounts for the fact that even if the distribution of matter would be homogeneously distributed, the number of discrete sources tracing this homogeneous background is not perfectly homogeneous, but it follows a Poisson distribution around the mean. This also impacts the PTA signal, see Ref.~\cite{Allen:2024mtn}, but is not subject of this work. Here, we focus on the galaxy clustering contribution, which constitutes an intrinsic source of anisotropy independent of shot noise.

\begin{table*}[!t]
  \centering
  \begin{adjustbox}{width=\textwidth}
  \normalsize
  \renewcommand{\arraystretch}{1.5}
		\begin{tabular}[c]{ p{0.25 \textwidth} p{0.03 \textwidth} p{0.55\textwidth} p{0.03 \textwidth}  p{0.15 \textwidth}  }
				\hline \hline     
				Standard variance && Variance due to the stochasticity of the GW emission, assuming an isotropic distribution of sources and one pulsar pair && Eq.~(C28) in \cite{Allen:2022dzg} \\
				 Clustering variance &&  Variance due to the stochasticity in the anisotropic distribution of GW sources induced by galaxy clustering, for one pulsar pair  && Eq.~\eqref{Eq:2ndmom_rho_ab_clustering}, \newline Eq.~(31) in \cite{Grimm:2024lfj} \\ \hline
				Pulsar-averaged standard \newline variance  (cosmic variance) &&  Standard variance in the continuous pulsar pair limit && Eq.~\eqref{Eq:irr_standard}, \newline Eq.~(C45) in \cite{Allen:2022dzg}  \\
    Pulsar-averaged clustering \newline variance && Clustering variance in the continuous pulsar pair limit && Eq.~\eqref{eq:irredclustvariance} \\
				\hline \hline
		\end{tabular}
  \end{adjustbox}
  \caption{\label{table_summary} We list the different contributions to the variance of the HD correlation, including a description and their defining equation. The first two lines correspond to the case where the signal arises from only one pulsar pair. The remaining lines correspond to the observational limit of a continuous sample of pulsars on the sky.} 
   \label{TableBig}
\end{table*}

\subsection{Galaxy clustering variance}\label{AnalyticResults}

In Ref.~\cite{Grimm:2024lfj}, we have derived an analytical expression for the contribution of large-scale galaxy clustering on the variance of the HD correlation, denoted by $\langle\rho_{ab}^2\rangle^\clust$.\footnote{Note again that there is no contribution of galaxy clustering to the mean of the HD correlation, which can be expressed as $\langle\rho_{ab}\rangle^\clust=0$.} This calculation is based on a two-step averaging procedure where we first assume a fixed galaxy density distribution and then take ensemble averages over all such distributions. The resulting variance is given by
\begin{align}
\angbr{\rho^2_{ab}}^\clust=&\iint\mathrm dz\,\mathrm dz'\left[G(z) G(z')+ \Gamma(z,z')\right]\iint\frac{\mathrm d\bn}{4\pi}\frac{\mathrm d\bn'}{4\pi} \xi_g\big(\bn\cdot\bn',z,z'\big)\chi_{ab}(\bn)\chi_{ab}(\bn') \nonumber \\ 
&+4 \iint\mathrm dz\,\mathrm dz'\,\Gamma(z,z') \iint\frac{\mathrm d\bn}{4\pi}\frac{\mathrm d\bn'}{4\pi} \xi_g\big(\bn\cdot\bn',z,z'\big)\chi_{aa}(\bn)\chi_{bb}(\bn')\,. \label{Eq:2ndmom_rho_ab_clustering}
\end{align}
Here, $\xi_g\big(\bn\cdot\bn',z,z'\big)$ denotes the galaxy correlation function
\be
\xi_g\big(\bn\cdot\bn',z,z'\big)\equiv \langle\delta_g(\bn,z)\delta_g(\bn',z')\rangle\,.
\ee
It encodes the correlation between overdensities of galaxies in two different directions $\bn$ and $\bn'$ of the sky (depending only on the angle between them due to statistical isotropy) and at redshifts $z$ and $z'$. The form and amplitude of this correlation function directly governs the clustering variance. 
The functions $G(z)$ and $\Gamma(z,z')$ in Eq.~\eqref{Eq:2ndmom_rho_ab_clustering} are measures of the GW strain weighted by $b_{\rm GW}(z)$, 
\begin{align}
G(z)&\equiv\int_0^\infty \mathrm df\,\bar S_h(f)b_{\rm GW}(z)\,, 
\nonumber \\
\Gamma(z,z')&\equiv\frac 14\int_0^\infty\int_0^\infty\mathrm df\,\mathrm df'\,\mbox{sinc}^2\rbr{\pi(f-f')T}
\bar S_h(f)\bar S_h(f')b_{\rm GW}(z)b_{\rm GW}(z')\,, \label{Eq:G_Gamma}
\end{align}
where $\text{sinc}(x)=\sin(x)/x$. We note that the frequency range of integration is theoretically infinite for a detector sensitive to arbitrary frequencies. In practice, however, the frequency bounds of integration should be chosen to correspond to the frequency band of the experiment considered. For our numerical evaluations, we choose a lower bound $f_{\rm min}=2.1\mathrm{nHz}$ (corresponding to $1/T$ for an observational time $T = 15\,\mathrm{yr}$, as in the latest NanoGrav data release~\cite{NANOGrav:2023gor}) and an upper bound $f_{\rm max}=100\,\mathrm{nHz}$. We note that the dependence of our results on the precise upper bound is small. Moreover, we note that in a realistic scenario, a sensitivity function should as well be included. Here, we aim to provide a theoretical prediction independently of the observational method, and thus omit any detector-specific sensitivity function.\footnote{We see from Eqs.~\eqref{Eq:HD_strain} and~\eqref{Eq:G_Gamma} that $G(z)$ and the squared strain $h^2$ have in fact the same frequency dependence. Thus, for the terms proportional to $G(z)$, the relative size of the variance compared to the signal will not be affected by a sensitivity function or a different choice of frequency bounds. The terms proportional to $\Gamma(z,z')$ are in principle affected by these details, although we do not expect them to significantly alter our results given their overall small amplitude (see Section~\ref{sec:results_results}).}

Eq.~\eqref{Eq:2ndmom_rho_ab_clustering} constitutes the main result of our previous work, Ref.~\cite{Grimm:2024lfj}. It corresponds to the case of observing a single pulsar pair. In Appendix~\ref{App:Shot_pulsar_variance}, we show the analogous derivations for the pulsar-averaged variances, assuming a continuous distribution of infinitely many pulsars on the sky. Note that similar derivations have recently been presented in Refs.~\cite{Allen:2024bnk, Agarwal:2024hlj}. To move from this theoretical framework to numerical evaluations, we need to specify an astrophysically realistic model for the function $b_{\rm GW}(z)$. This is done in the following section, applying three different models of SMBH mergers.

\section{Astrophysical modeling} \label{sec:ast_modeling}

The clustering variance depends on the mass and redshift distribution of sources and on the astrophysical kernel $b_{\rm GW}(z)$, defined through Eq.~\eqref{eq:sh_clust}. Moreover, the mean spectral density $\bar{S}_h$ affects the normalization of the HD correlation and its variance. We now model these quantities considering different astrophysical models for SMBH formation and evolution.

\subsection{Modeling of $b_{\rm GW}$} \label{Sec:AstroTh}

In order to derive an expression for the astrophysical kernel $b_{\rm GW}(z)$ we follow different steps: first, we derive an expression for the background energy density, in terms of source distribution and emission rate. Then we relate energy density to spectral density $S_h$, and we determine the kernel by comparing with the parametrization in Eq.~\eqref{eq:sh_clust}.

Any GW carries a certain energy flux $\Phi$ (flow  of energy per unit time per unit area), and the local flux can be related to the gauge invariant gravitational energy density $\rho_\text{GW}$ by
\cite{Maggiore:1900zz}
\be
\rho_\text{GW} = c^{-1} \Phi\,.
\ee
The flux associated with an astrophysical SGWB is the cumulative flux of all astrophysical sources.
Thus, to each comoving volume element $\dd V_c$ located at comoving distance $r_z$, we may assign a certain GW luminosity $L_\text{GW}$, corresponding to the GW energy released per unit source time within this volume, see Refs.~\cite{Cusin:2017fwz, Dvorkin:2016okx, Phinney} for more details. The energy density (or flux) is then the integral over comoving distance $r_z$ of the luminosity $L_\text{GW}$, with an inverse square luminosity distance factor $d_L^{-2}$ accounting for the fact that source frequencies and the energies are redshifted and spread out onto a sphere\footnote{Note that the redshift $z$ and the luminosity distance $d_L$ in Eq.~\eqref{eq:rhoIntegral} also depend on $\bn$, since in an anisotropic universe, these quantities are not only due to the expansion of the Universe, but also to Doppler effects, gravitational redshift and integrated Sachs-Wolfe, see e.g.\ Ref.~\cite{Bonvin:2011bg}. The anisotropy due to the source distribution dominates however over the redshift and luminosity distance anisotropies and we neglect the latter here.}, that is:
\be
\rho_\text{GW}(\bn) = c^{-1} \int \dd r_z \, r_z^2 \ \frac{1}{4\pi d_L^2} \frac{\dd L_\text{GW}}{\dd V_c} = \frac{1}{4\pi} \int \dd z \ \frac{1}{(1+z)^2 H(z)} \frac{\dd L_\text{GW}}{\dd V_c}(\bn, z)\,,\label{eq:rhoIntegral}
\ee
where we introduced the Hubble parameter $H(z)$ by changing variable from $r_z$ to $z$. Since GW sources are not isotropically distributed on the sky, the luminosity $L_{\rm GW}$ and thus the energy density $\rho_{\rm GW}$ are functions of the sky direction $\bn$.

The luminosity per comoving volume element in Eq.~\eqref{eq:rhoIntegral} can be expressed as the integral over all masses and frequencies of the individual power $\dd E_s/\dd t_s$ radiated by each source within this volume element
\begin{align}
    \frac{\dd L_\text{GW}}{\dd V_c}  =& \iint \dd \log(\M) \,\dd f_s \ \frac{\dd N_s}{\dd V_c \dd \M \dd f_s} (\bn, z, \M, f_s ) \ \frac{\dd E_s}{ \dd t_s} (\log(\M), f_s) \\ =& \iint \dd \log(\M)\,\dd f \ \left.\left(\frac{\dd N_s}{\dd V_c \dd \log(\M) \dd f} \ \frac{\dd E_s}{ \dd t_s}\right)\right|_{ (\bn, z, \M, f_s = f (1+z))} \,, \label{Lshell}
\end{align}
where $\mathcal{M}$ is the chirp mass of the source given by
\be
\label{eq:chirp}
\mathcal{M}= \frac{(m_1 m_2)^{3/5}}{(m_1+m_2)^{1/5}}\,,
\ee
with $m_{1,2}$ the masses of the two BHs in the binary. $N_s$ denotes the number of sources, $E_s$ is the source-frame energy of one source with mass $\M$ and (source-frame) frequency $f_s$, and $t_s$ is the local cosmic time at the source. In Eq.~\eqref{Lshell}, we changed variable from the source-frame frequency $f_s$ to the observed frequency $f = f_s/(1+z)$. 

Denoting the number of sources per comoving volume $\dd V_c$, per frequency $\dd f$ and per logarithmic chirp mass $\dd\log{\mathcal{M}}$ by
\be
n_s(\bn, z, \M, f ) \equiv \frac{\dd N_s}{\dd V_c\, \dd \log(\M)\, \dd f}\,,
\ee
we then split it into an angle-averaged term $\bar{n}_s$ and an anisotropic contribution
\be
n_s(\bn, z,\mathcal M, f)=\bar{n}_s(z,\mathcal M, f)+\delta n_s(\bn, z,\mathcal M, f)\, .
\ee
We introduce the energy density parameter $\Omega_{\rm GW}$, 
\be
\Omega_{\rm GW} \equiv \frac{1}{\rho_c}\frac{\dd \rho_{\rm GW}}{\dd \log f}\,,
\ee
as the ratio of the energy density of the astrophysical SGWB per logarithmic observed frequency, $\dd \rho_{\rm GW}/\dd \log f$, to the critical energy density $\rho_c$ of the Universe.
 To consider the contributions of the different directions $\bn$ to the total energy density parameter at the observer, we then analogously split $\Omega_{\rm GW}$ in an angle-averaged term $\bar\Omega_{\rm GW}$ and an anisotropic fluctuation, as 
\be
\Omega_{\rm GW}(f, \bn)=\bar{\Omega}_{\rm GW}(f)+\delta\Omega_{\rm GW}(f, \bn)\,.\label{eq:Omega_splitting2}
\ee
We now focus on the clustering contribution $\delta n_s$ to the anisotropies in the distribution of sources in order to extract the quantity $b_{\rm GW}$. Denoting the fractional fluctuation in the number density of sources by 
\be
\delta_s(\bn, z, \mathcal{M})=\frac{\delta n_s(\bn, z, \mathcal{M}, f)}{\bar{n}_s(z, \mathcal{M}, f)}\,,
\label{eq:deltas}
\ee
and using that the GW sources are biased tracers of the underlying distribution of galaxies, 
\be
\delta_s(\bn, z, \mathcal{M})=b_{s}(z, \mathcal{M})\delta_g(\bn, z)\, , \label{eq:def_bias}
\ee
we obtain 
\begin{equation}
\delta\Omega_{\rm GW}(f, \bn) = \frac{f}{4\pi\rho_c}\iint \dd z\,\dd \log(\M)  \ \frac{1}{(1+z)^2 H(z)}\, \frac{\dd E_s}{\dd t_s} \ \bar n_s(z,\M, f) b_s(z,\M) \, \delta_g(\bn, z)\,.
\label{eq:deltaOmega}
\end{equation}
Eq.~\eqref{eq:deltaOmega} relates the anisotropies in the energy density of an astrophysical SGWB to the anisotropies of the large-scale structure where SGWB sources are found. The bias $b_s(z,\M)$, defined through Eq.~\eqref{eq:def_bias}, relates the fractional galaxy anisotropies to the fractional source anisotropies. It does not depend on direction, since we assume that GW sources follow the same clustering structure as the galaxies, but it can depend on mass and redshift. Note that in Eq. \eqref{eq:deltas}, we assumed that given a mass and a redshift, the clustering contribution shares the same frequency distribution as the background, so that $\delta_s$ does not depend on $f$. In other words, a source with parameters $(z, \M)$ traces the galaxy identically no matter at what frequency it is emitting.

The spectral density of a SGWB given in Eq.~\eqref{eq:sh_clust} is directly related to its energy content via the relation \cite{Maggiore:2018sht} 
\be
\Omega_{\rm GW} = c^2 f^3 S_h/(8G\rho_c)\,. 
\label{eq:OmegaSh_relation}
\ee
In particular, for the anisotropic part, we arrive at 
\be
\delta S_h =  \delta\Omega_{\rm GW}\  \frac{8 G\rho_c}{ f^3 c^2}=  \bar S_h(f)\int\mathrm d z\, b_{\rm GW}(z)\delta_g(\mathbf{n},z) \,, \label{Eq:delta_sH}
\ee
where we have used Eq.~\eqref{eq:sh_clust} for the definition of $\delta S_h$. The form of the gravitational energy released by astrophysical sources was so far generic. We now focus on the case of an inspiraling BBH, for which we have \cite{Maggiore:1900zz} 
\be
\frac{\dd E_s}{\dd t_s}  =\frac{96}{15} \pi^{10/3}\frac{1}{Gc^5} (G\mathcal{M})^{10/3} f_s^{10/3} =\frac{96}{15} \pi^{10/3}\frac{1}{Gc^5} (G\mathcal{M})^{10/3} (1+z)^{10/3} f^{10/3}\,. \label{eq:dEdlogf}
\ee
Combining Eqs.~\eqref{Eq:delta_sH},~\eqref{eq:dEdlogf} and~\eqref{eq:deltaOmega} we obtain
\be
b_{\rm GW}(z) = \frac{1}{\bar S_h(f)}\int\dd\log(\mathcal{M}) b_s(z,\M)\alpha(z,\M,f)\,, \label{Eq:astro_result}
\ee
where 
\be
\alpha(z,\M, f)\equiv\frac{64}{5} \frac{\pi^{7/3}}{c^7 H(z)} (G \M)^{10/3} f^{4/3} (1+z)^{4/3}\ \bar n_s(z, \M, f)\, . \label{Eq:alpha}
\ee
 Eq.~\eqref{Eq:astro_result} provides an expression for the gravitational kernel $b_{\rm GW}(z)$ in terms of the distribution of sources $\bar{n}_s(z,\M, f)$, the mean spectral density $\bar{S}_h(f)$ and the source bias $b_s(z,\M)$. We have so far allowed the $f$-dependency in the distribution of sources $\bar n_s$ to be generic. However, note that in the scenario of a distribution of supermassive binaries evolving via quasi-circular inspirals, the frequency distribution is, in the limit of many very massive sources \cite{Sesana:2008mz},
\be
\bar n_s(z, \M, f) = \bar n_s\left(z, \M, f = 1 \text{Hz}\right) \left(\frac{f}{1 \text{Hz}}\right)^{-11/3}\,,
\ee
corresponding to the idea that the probability for a binary to be found in a frequency interval $\Delta f$ is proportional to the time it spends in the $\Delta f$ interval. In this case, we see that $b_{\rm GW}(z)$ does not depend on frequency $f$, since in Eq.~\eqref{Eq:astro_result} both $\alpha(z,\M, f)$ and $\bar S_h(f)$ scale as $f^{-7/3}$. As a consequence, $\bar S_h(f)$ and $\delta S_h(f)$ have the same frequency dependence, see Eq.~\eqref{eq:sh_clust}.\footnote{ Even in scenarios where $\bar S_h(f)$ does not scale as $f^{-7/3}$, i.e.~where the GW background is not perfectly described by quasi-circular inspiraling SMBH mergers, we generally expect $\bar S_h(f)$ and $\delta S_h(f)$ to scale equally with frequency and thus $b_{\rm GW}(z)$ to be frequency-independent. A frequency dependence of $b_{\rm GW}(z)$ could only arise if regions with higher source density would exhibit a different frequency scaling as regions with lower source density.}
 
We can extract the sky-averaged quantities by repeating the same calculation and substituting the quantity $\bar n_s(z,\M, f) b_s(z,\M) \, \delta_g(\bn, z)$ with the sky-averaged quantity $\bar n_s(z,\M, f)$ in Eq.~\eqref{eq:deltaOmega}. This leads to
\be
\label{eq:barS}
\bar S_h(f)=\int\mathrm dz\int\mathrm d\log(\M)\,\alpha(z,\M,f)\,.
\ee

\subsection{Reconstruction of $b_{\rm GW}$ from catalogs} \label{sec:catalogs}

From Eqs.~\eqref{Eq:astro_result} and \eqref{Eq:alpha}, the contribution of source anisotropies to the variance of the HD correlation must rely on a model for the (isotropic) distribution in chirp mass and volume of inspiraling SMBHs, $\bar n_s(z,\M, f)$. This distribution follows from the distribution of individual SMBHs, combined with some modeling of the binary formation process.
The former has observational constraints from electromagnetic signals of high redshift SMBH-powered quasars, other active galactic nuclei emission processes, as well as baryon dynamics, see Refs.~\cite{Sargant_historicalWeighting, Blanford_ReverberationMapping, Peterson_ReverberationMapping, Fan1, Ferrarese_Msigma} for a selection of historical approaches. The theoretical modeling of this SMBH population is however a long-term astrophysical challenge closely tied to the mechanism via which such SMBH may arise, see Ref.~\cite{inayoshiReview, volonteriReview} for reviews. In particular, from naive models it is hard to predict the observations of heavy SMBH at redshifts as early as $z=6$ \cite{Turner1991quasars}.

The merging process is a further challenge due to the wide range of scales involved, ranging from Mpc scale for galaxy halo mergers to $10^{-6}$pc for the subsequent SMBH merger. Owing to this difficulty, we rely here on the numerical catalogs of SMBH binaries described in Refs.~\cat. These catalogs explore two different scenarios for the production of seeds of individual SMBH:
\begin{enumerate}
    \item heavy ($\sim 10^5 M_\odot$) seeds following the collapse of protogalactic disks at $10\lesssim z\lesssim 15$,
    \item light ($\sim 150 M_\odot$) seeds following the collapse of population III stars at $z\sim 20$.
\end{enumerate}
These seeds are then evolved within a model for galaxies that includes prescriptions for the evolution and mergers of dark matter halos, baryonic gas dynamics, and environmental effects, thereby providing a description of the co-evolution of SMBHs and their galactic hosts. We refer to Refs.~\cat\ for more details on the catalog implementation. 

For the case of seeds from the collapse of protogalactic disk, the catalog exists in two variants. In one case, a non-zero time delay $\Delta z_\text{del}$ between galaxy and SMBH merger is implemented, while in the other it is explicitly set to $\Delta z_\text{del}=0$. These variants are named {\tt Q3-d} and {\tt Q3-nd}, with {\tt Q3} referring to the value of the Toomre stability parameter $Q=3$ of the collapsing protogalactic disk, while {\tt d} and {\tt nd} respectively refer to the assumption of a delay or no delay. For the case of population III star seeds, the assumptions on the delay $\Delta z_\text{del}$ have been found to not appreciably alter the population of sources, hence the population III seeds model is only provided in one variant, labeled as {\tt popIII}.

\begin{figure}
\centering
    \includegraphics[width = 0.5 \textwidth]{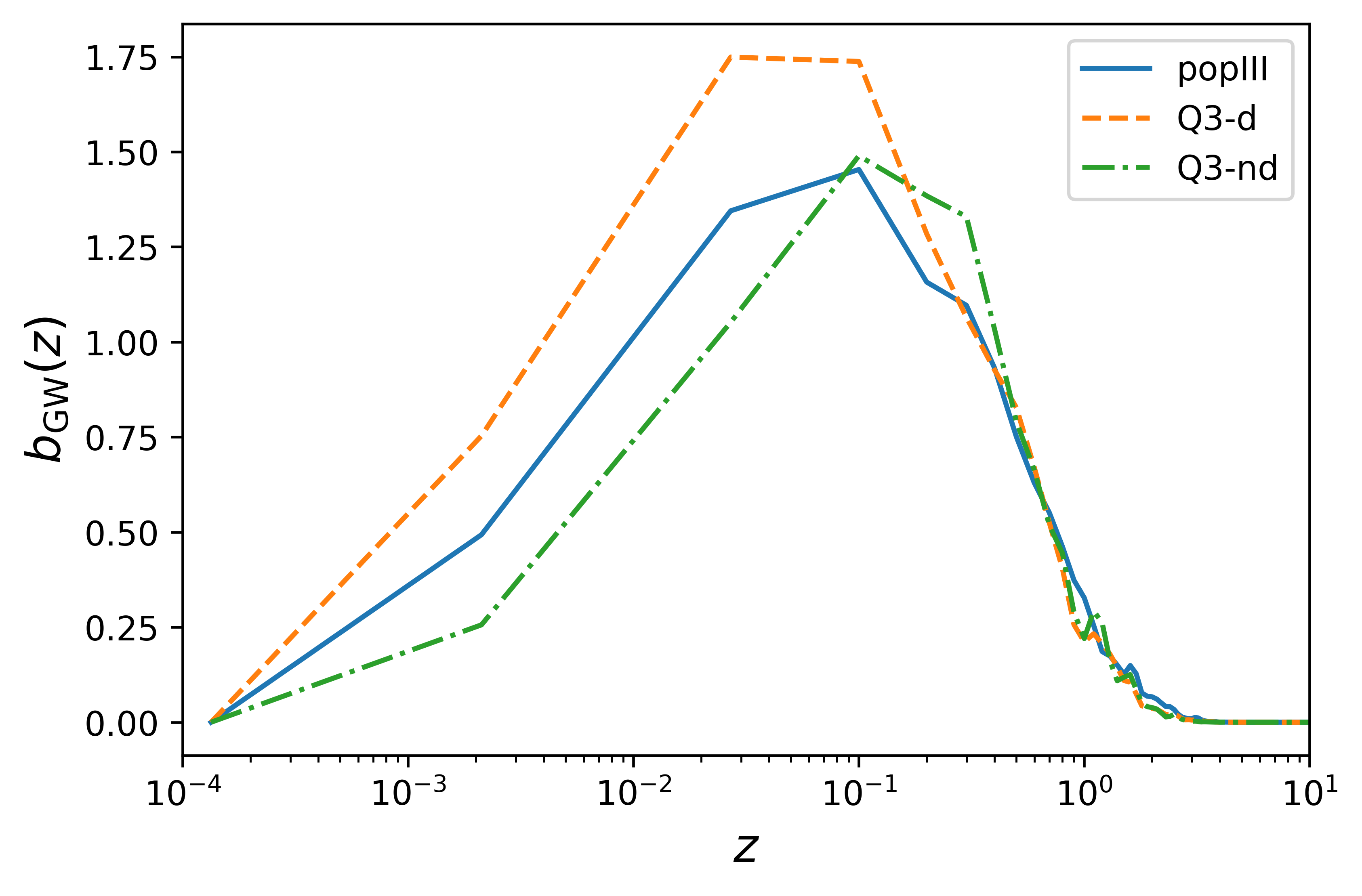} 
    \caption{The function $b_{\rm GW}(z)$ is shown for the three different population models in consideration.} 
    \label{fig:bGW}
\end{figure}

In Appendix \ref{app:numerics}, we outline how to use these numerical catalogs of SMBH binaries to reconstruct the isotropic GW luminosity over $z$, assuming that binaries follow quasi-circular orbits. We then extract from it a numerical expression of the astrophysical kernel $b_{\rm GW}$ of Eq.~\eqref{Eq:astro_result}. We plot our results in Fig.~\ref{fig:bGW} for the three catalog models, which all show a similar behavior for the astrophysical kernel. From Fig.~\ref{fig:bGW}, we see that the astrophysical signal to which PTA are sensitive arises from sources at $z\lesssim 2$, with a peak at $z\sim0.1$.  The {\tt popIII} model produces a catalog of lighter but more numerous binary sources compared to the {\tt Q3} models, such that, after performing the integral in Eq.~\eqref{Eq:astro_result}, their astrophysical kernels become comparable. This is true whether or not one assumes $\Delta z_\text{del} =0$.

\subsection{Discussion on the source bias and galaxy bias} \label{sec:bias}

In our numerical computation of $b_{\rm GW}$ from Eq.~\eqref{Eq:astro_result} we have set the source bias to $b_s(z,\M)=1$. This source bias encodes a) the way the distribution of GW sources (inspiraling SMBH in our case) traces the distribution of galaxy mergers, and b) the way the distribution of galaxy mergers traces the distribution of galaxies. The merger rate of two galaxies, resulting from the merger of the two host dark matter halos, is actually mass dependent as can be predicted e.g.~with the extended Press-Schechter formalism~\cite{LaceyCole}. More specifically, it is higher for higher mass galaxies and halos, and in principle depends on redshift as well. From the Millennium numerical simulations, this dependency has been found to be of the type $\sim (1+z)^{0.1} M_h^{0.13}$, with $M_h$ being the mass of the merged halo~\cite{fakhouri2010merger}. Given this weak dependency, one may simplify the description by assuming that galaxy mergers trace the galaxy population identically at all masses and redshifts. 

Using this simplification, the galaxy fractional number overdensity $\delta_g$ then equals the fractional overdensity of galaxy merger number counts, as no significant bias arises from b), i.e.\ the way the distribution of galaxy mergers traces the distribution of galaxies. We further note that the overdensity of galaxy mergers counts corresponds to the fractional number overdensity of inspiraling SMBH $\delta^{\rm clust.}_s$, with a possible delay $\Delta z_\text{del}$ owing to the typical time delay between galaxy merger and SMBH merger, such that $\delta^{\rm clust.}_s(\bn, z) \simeq \delta_g(\bn, z+\Delta z_\text{del})$. Setting $\Delta z_\text{del} = 0$, assuming that the time span of such mergers is small compared to the cosmological time span over which the galaxy overdensity evolves, we arrive at $\delta^{\rm clust.}_s(\bn, z) \simeq \delta_g(\bn, z)$. In other words, also no significant bias arises from a), i.e.\ the way the distribution of galaxy mergers traces the distribution of galaxies, and we arrive at $b_s(z,\M)\simeq 1$. The fact that, as seen above, catalogs with and without delay $\Delta z_\text{del}$ lead to similar values of the mass integral over $\alpha$ indicates that this approximation is reasonable.

Another bias factor that also affects our numerical results is the galaxy bias $b_g(z)$. Indeed, the quantity that we use in Eq.~\eqref{Eq:2ndmom_rho_ab_clustering} to describe the impact of galaxy clustering on the HD correlation is the galaxy correlation function $\xi_g(\bn\cdot\bn',z,z')$. However, cosmological models instead make prediction on the matter correlation function, $\xi_m(\bn\cdot\bn',z,z')$. Since galaxy over- or underdensities are not perfect tracers of the underlying matter field, these quantities are related via the galaxy bias:
\be
\xi_g(\bn\cdot\bn',z,z')=b_g^2(z)\,\xi_m(\bn\cdot\bn',z,z')\,.
\ee
The galaxy bias itself can be expressed as a number-weighted integral over the halo bias $b_h(M_h)$, depending on halo mass $M_h$~\cite{Desjacques:2016bnm}. Thus, our astrophysical modeling would in principle need to be extended by choosing a model for the halo bias (see e.g.\ Fig.~15 in Ref.~\cite{Desjacques:2016bnm}) and relating halo masses to central BH masses as outlined in Ref.~\cite{Ferrarese:2002ct}. This bias, depending on BH mass, could then be included in Eq.~\eqref{Eq:astro_result} for $b_{\rm GW}$, and the integral over chirp mass $\M$ would be replaced by integrals over the individual BH masses available from the catalogs~\cat.  
However, as we can infer from Fig.~15 in Ref.~\cite{Desjacques:2016bnm}, halo bias reaches a maximum value of $4-5$ for the most massive halos only, and thus we expect the overall boost of our signal due to this bias to be below this value. In the following, we set the galaxy bias to 1 for simplicity. As we will see, since the clustering variance is roughly two orders of magnitude smaller than the standard variance, even a boost of a factor 5 would not make this contribution relevant for observations.

\section{Numerical results for the clustering variance} \label{sec:results}

Having established our astrophysical modeling, we are now equipped to perform numerical evaluations for the galaxy clustering variance of the HD correlation. In the following, we will first describe our numerical methods and then present our results. A comparison to Ref.~\cite{Allen:2024mtn}, a recent work with similar considerations, is provided as well.

\subsection{Methods}

The galaxy clustering variance in Eq.~\eqref{Eq:2ndmom_rho_ab_clustering} {poses considerable challenges for numerical evaluation}, since it contains a 6 dimensional integral (four over angles and two over redshifts) of the galaxy correlation function $\xi_g(\bn\cdot\bn',z,z')$, weighted by the highly non-trivial functions $\chi_{ab}(\bn)$. In Ref.~\cite{Grimm:2024lfj}, we presented alternative expressions using either the galaxy power spectrum $P_g(k,z,z')$ or the angular power spectrum $C_l(z,z')$ defined via
\begin{equation}
\angbr{\delta_g(\mathbf{k},z)\delta_g^\ast(\mathbf{k}',z')}=(2\pi)^3\delta_D(\mathbf{k}-\mathbf{k'})P_g(k,z,z')\,, \qquad \xi_g(\bn\cdot\bn',z,z')=\frac{1}{4\pi}\sum_l (2l+1)C_l(z,z')\mathcal{P}_l(\bn\cdot\bn')\,, \label{Eq:Cell}
\end{equation}
where $\delta_g(\mathbf{k},z)$ is the Fourier transform of the galaxy overdensity and $P_l$ represents the $l$-th Legendre polynomial. Indeed, the integral over $\bn$ and $\bn'$ in the second line of Eq.~\eqref{Eq:2ndmom_rho_ab_clustering} has a convenient analytical solution when using the angular power spectrum $C_l(z,z')$. We find that
\begin{equation}
    \iint\frac{\mathrm d\bn}{4\pi}\frac{\mathrm d\bn'}{4\pi}\mathcal{P}_l\rbr{\bn\cdot\bn'} \chi_{aa}(\bn)\chi_{bb}(\bn')=\frac{1}{9}\rbr{\delta_{l0}+\frac{1}{4}\cos\gamma\,\delta_{l1}+\frac{1}{400}\rbr{1+3\cos(2\gamma)}\delta_{l2}}\,, \label{Eq:Integral2}
\end{equation}
where $\gamma$ denotes the angle between the two pulsars at $a$ and $b$.\footnote{This expression can be shown by first choosing sky coordinates such that the pulsar $a$ aligns with the $z$-direction, $\mathbf{n}_a=\mathbf{e}_z$, and the pulsar $b$ lies on the $x-z$ axis, $\mathbf{n}_b=\sin\vartheta\mathbf{e}_x+\cos\vartheta\mathbf{e}_z$. Then, the quantities $\rho_{aa}(\bn)$ and $\rho_{bb}(\bn')$ as well the Legendre polynomial $\mathcal P_l(\bn\cdot\bn')$ can be decomposed into spherical harmonics (similar in spirit to Ref.~\cite{Allen:2024bnk}), and applying the orthogonality of spherical harmonics finally leads to Eq.~\eqref{Eq:Integral2}.}
As a consequence, the second line in Eq.~\eqref{Eq:2ndmom_rho_ab_clustering} only contributes to the variance if a monopole ($l=0$), dipole ($l=1$) or quadrupole ($l=2$) in the galaxy overdensities would be present. However, the monopole is set to zero by construction (any non-zero monopole can be reabsorbed in the strain and thus in the normalization of the HD curve), while contributions at ultra-large angular scales $l=1,2$ are small unless some mechanism breaking the cosmological principle is present in the universe.\footnote{Some observations indicate the presence of structure at ultra large angular scales, such as the cosmic dipole that has been measured from the overabundance of radio surveys~\cite{Colin:2017juj,Bengaly_2018,Secrest:2020has,Siewert:2020krp,Secrest:2022uvx}. However, this anomaly might be due to systematic effects and independent measurements of the cosmic dipole, e.g.~with ground based GW observatories~\cite{Grimm:2023tfl, Mastrogiovanni:2022nya, Cousins:2024bhk}, could bring more clarity.} Here, we stay within the scope of the standard $\Lambda$CDM cosmological model obeying the cosmological principle. This means that the second line of Eq.~\eqref{Eq:2ndmom_rho_ab_clustering} cannot lead to any sizable contribution, and we thus focus on the first line. 

In the following, we numerically evaluate the remaining term in Eq.~\eqref{Eq:2ndmom_rho_ab_clustering}, having contributions from all galaxy clustering multipoles. To compute this term numerically, it is more convenient to rewrite the correlation function in terms of the power spectrum $P_g(k,z)$. We can then apply the Limber approximation~\cite{limber1953analysis, Bartelmann:1999yn} to reduce from two to one integral over redshift,
\begin{align}
\angbr{\rho^2_{ab}}^\clust=&\int\mathrm dz\left[{G^2(z)}+\Gamma(z,z)\right]\frac{H(z)}{c}\iint\frac{\mathrm d\bn}{4\pi}\frac{\mathrm d\bn'}{4\pi}\chi_{ab}(\bn)\chi_{ab}(\bn') \int\frac{k\,\mathrm dk}{2\pi}P_g(k,z) J_0\big(k r_z\cos^{-1}(\bn\cdot\bn')\big)\,. \label{Eq:2ndmom_Limber}
\end{align}
Here, $J_0$ is the 0th order Bessel function, $H(z)$ is the Hubble parameter at redshift $z$, and $r_z$ denotes the comoving distance given by $r_z=\int_0^z\dd z'cH(z')^{-1}$. The Limber approximation has been shown to work well for galaxy clustering at small separation and in the case of wide redshift bins~\cite{LoVerde:2008re}. Moreover, it is more precise at low than at high redshifts~\cite{LoVerde:2008re, Martinelli:2021ahc}. Since the gravitational kernel $b_{\rm GW}$ has support up to redshift $z\sim 2$ and peaks at a low redshift $z\sim 0.1$ (see Fig.~\ref{fig:bGW}), we expect the Limber approximation to have a high precision. Only for very large angular separations, this approximation will not be accurate. However, at those large scales the amplitude of the galaxy correlations is very small, and we expect no significant contribution to the variance from such less accurate modes.

\begin{figure}
    \centering
    \includegraphics[width=.495\textwidth]{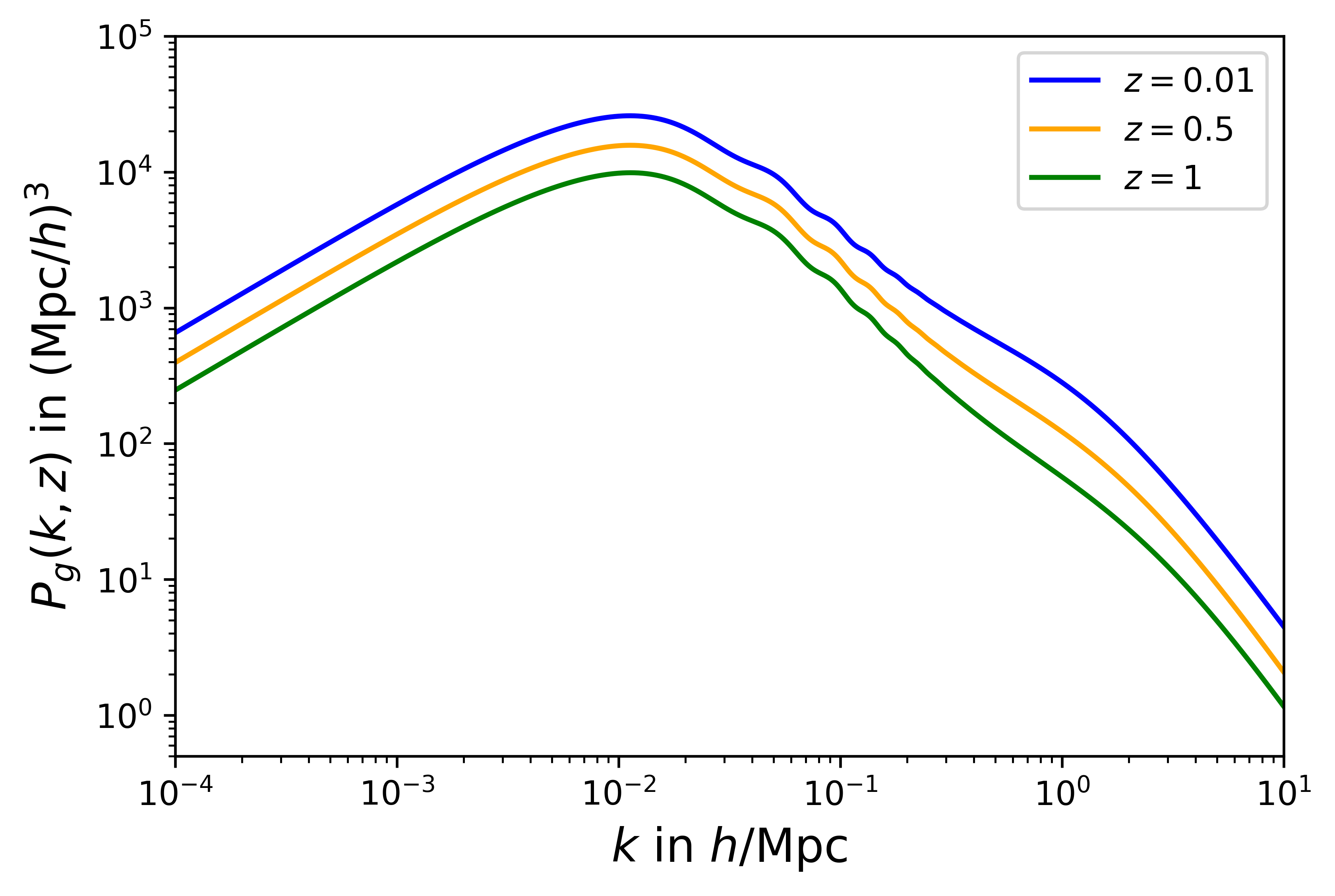}
    \includegraphics[width=.495\textwidth]{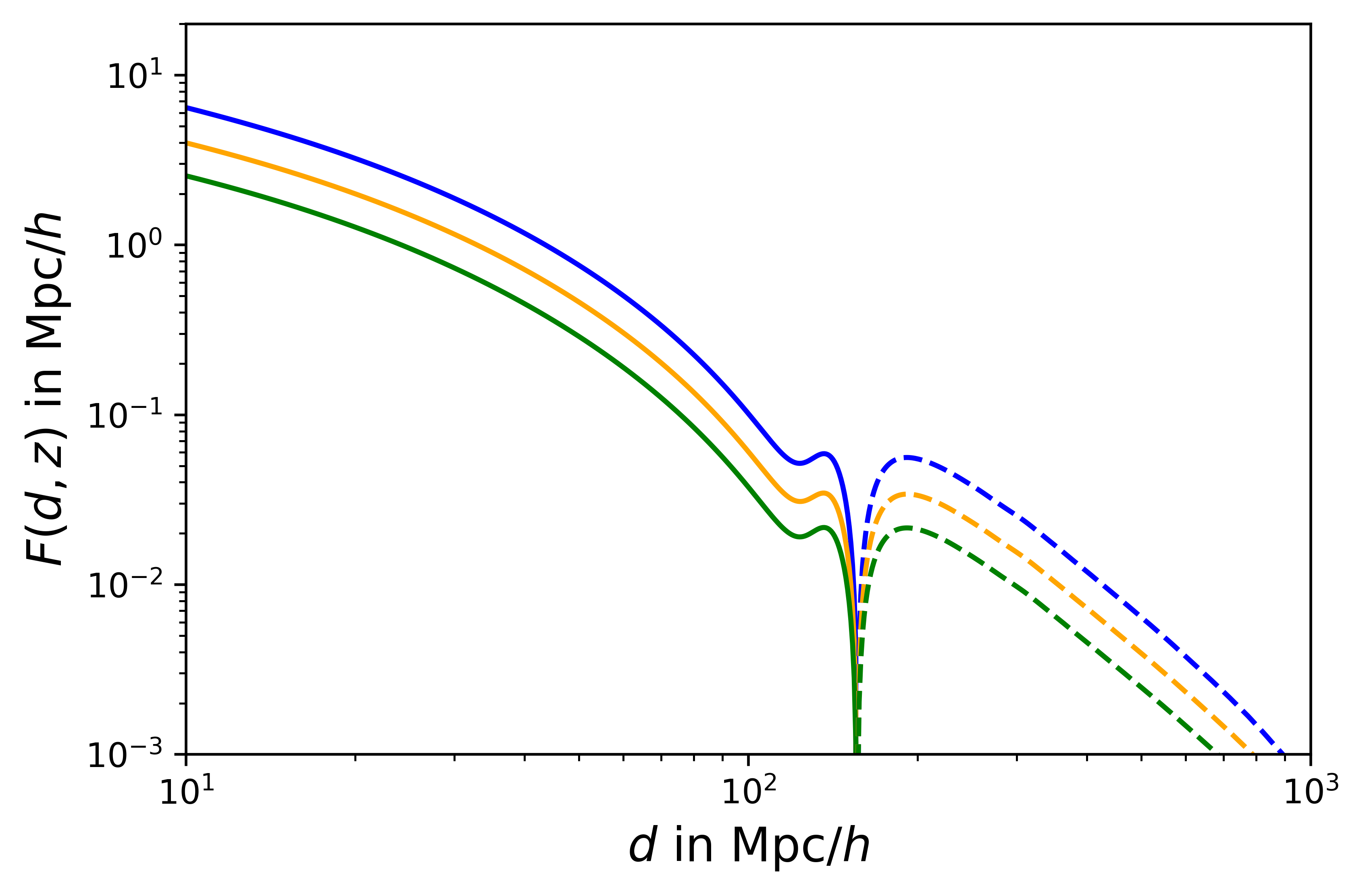}
    \caption{\textit{Left panel:} Galaxy power spectrum $P_g(k,z)$ (set equal to the matter power spectrum, ignoring galaxy bias) as a function of $k$ for 3 redshift values $z=0.01,0.5,1$. \textit{Right panel:} The function $F(d,z)$ at the same 3 redshifts. Dashed lines denote negative values.} \label{fig:Pk_F}
\end{figure}

\begin{figure}
    
    \centering
    \includegraphics[width = 0.6\textwidth]{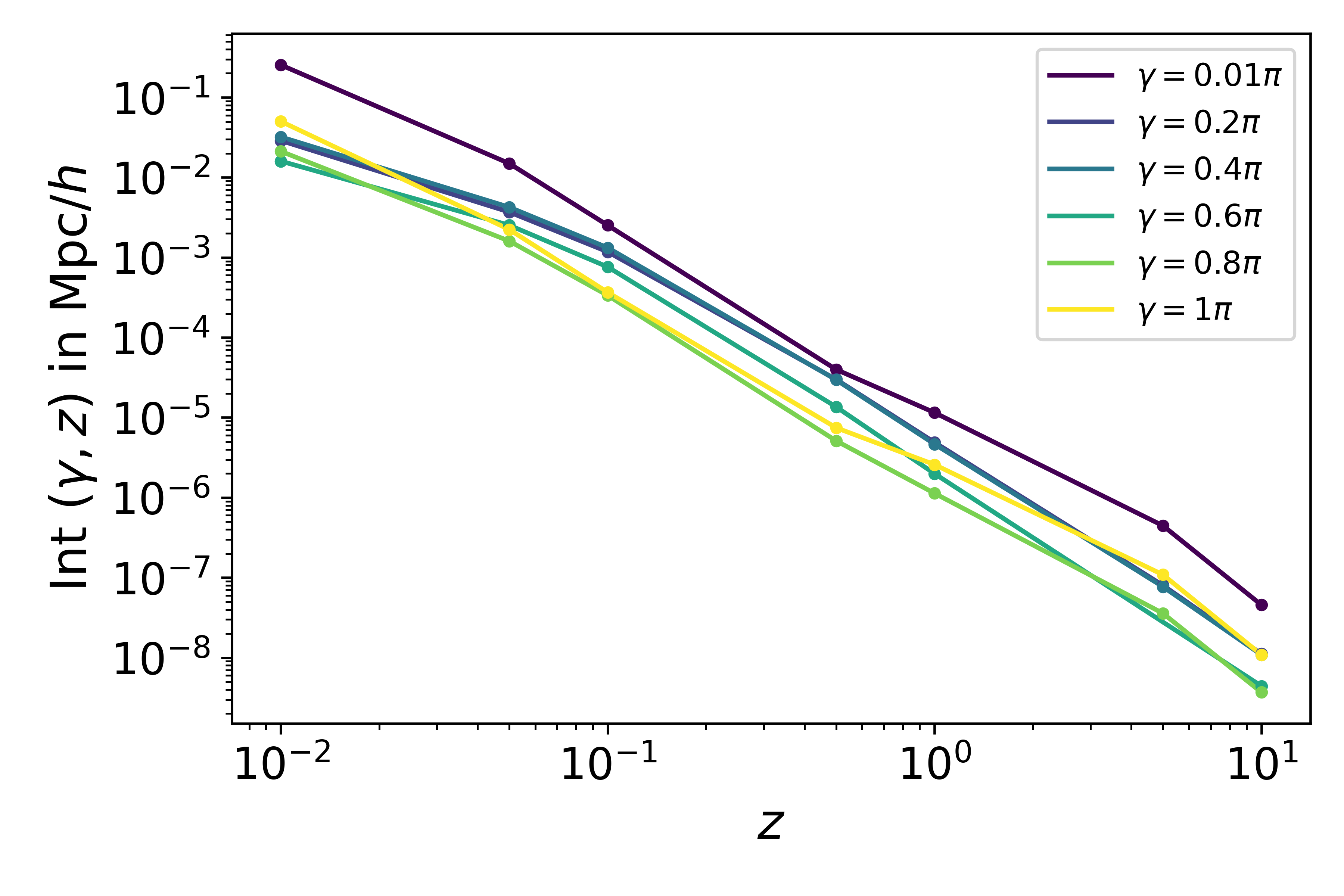} 
    \caption{Results for $\mathrm{Int}(\gamma,z)$ defined in Eq.~\eqref{Eq:Int_gamma_z} obtained from Monte Carlo integration at various pulsar pair separations $\gamma$, plotted as a function of redshift. For better comprehensibility of this figure, we omit the pulsar pair separations $\{0.1,0.3,0.5,0.7,0.9\}\pi$ and the redshift $z=0.0001$, where no unexpected behavior has been found.} 
    \label{fig:Int_gamma_z}

\end{figure}

Our numerical evaluation is performed using the following steps:
\begin{enumerate}
    \item First, we determine the galaxy power spectrum $P_g(k,z)$ at redshifts $\{0,0001,0.01,0.05,0.1,0.5,1,5,10\}$ by using the publicly available\footnote{\href{https://github.com/lesgourg/class_public}{https://github.com/lesgourg/class\_public}} 
    \texttt{CLASS} code~\cite{Lesgourgues:2011re} to obtain the matter power spectrum\footnote{The \texttt{CLASS} code provides the options to use either the linear or non-linear matter power spectrum. Here, we choose the non-linear option, as it is known to work well for high-mass halos~\cite{Zu:2012am}, corresponding to high-mass central BHs which contribute more to the GW signal. Another possible approach would be to use the linear power spectrum and then apply a non-linear halo bias~\cite{Mead:2020qdk}, although we choose to avoid such complications given the overall small amplitude of the clustering variance.}, and ignoring the bias $b_g$ between matter and galaxy correlations as discussed above. We assume cosmological parameters measured from the Planck survey~\cite{Planck:2018vyg}. 
     While we consider a limited amount of redshifts, we will see below that this choice is sufficient to describe the redshift-behavior of the integrals over angles $\bn$ and $\bn'$, which appear roughly linear on a double-logarithmic scale. The galaxy power spectrum $P_g(k,z)$, as a function of $k$ and for three values of $z$, is plotted on the left panel of Fig.~\ref{fig:Pk_F}.\footnote{The power spectrum is typically given in units of $(\mathrm{Mpc}/h)^3$, where $h$ (unrelated to the GW strain) sets the value of the Hubble parameter today, $H_0=100h\,\mathrm{km/s/Mpc}$. Our final results do not depend on $h$, as this factor cancels out in Eq.~\eqref{Eq:2ndmom_Limber}.}  
    \item The integral over $k$ in Eq.~\eqref{Eq:2ndmom_Limber} to obtain the function
    \be
     F(d,z)=\int\frac{k\,\mathrm dk}{2\pi}P_g(k,z) J_0\big(k d\big)\,,
    \ee
    is then performed using an FFTLog code.\footnote{ \href{https://github.com/JCGoran/fftlog-python}{https://github.com/JCGoran/fftlog-python}} We show the result, as a function of $d$ for three values of $z$, in the right panel of Fig.~\ref{fig:Pk_F}.
    \item Then, we numerically evaluate the integral over angles $\bn$ and $\bn'$,
    \begin{equation}
    \mathrm{Int}(\gamma, z)=\iint\frac{\mathrm d\bn}{4\pi}\frac{\mathrm d\bn'}{4\pi}\chi_{ab}(\bn)\chi_{ab}(\bn') F\big(r_z\cos^{-1}(\bn\cdot\bn'),z\big)\,,  \label{Eq:Int_gamma_z} 
    \end{equation}
    for values $\{0.01,0.1,0.2,0.3,0.4,0.5,0.6,0.7,0.8,0.9,1\}\pi$ of separation $\gamma$ between the pulsars $a$ and $b$. We perform a Monte Carlo integration $1000$ times using $10^7$ sampling points each, giving us both a mean and an error estimate (well within the marker size in Fig.~\ref{fig:Int_gamma_z}).\footnote{For each value of $\gamma$ and $z$, this computation needs roughly 3 hours runtime with the usage of 10 CPUs on the Yggdrasil cluster of the University of Geneva.} The results $\mathrm{Int}(\gamma, z)$ are plotted in Fig.~\ref{fig:Int_gamma_z} as a function of $z$ for the various values of $\gamma$.  We see that the functions exhibit a smooth redshift behavior. Therefore, no increase in the amount of redshifts is needed to obtain a reliable estimate of the galaxy clustering variance. We also note that these functions drop off fast with increasing $z$. Thus, we will find that $\sim 95\%$ of the clustering contribution to the variance arises from redshifts $z\leq 0.1$, where the astrophysical kernel $b_{\rm GW}(z)$ roughly peaks.
    \item Finally, to perform the integral over redshift in Eq.~\eqref{Eq:2ndmom_Limber}, we need to evaluate the kernels $G(z)^2$ and $\Gamma(z,z)$. These depend on the function $b_{\rm GW}(z)$ via Eqs.~\eqref{Eq:G_Gamma}. After obtaining $b_{\rm GW}(z)$ from the catalogs as described in Section~\ref{sec:catalogs}, we can combine with the redshift-interpolated functions $\mathrm{Int}(\gamma,z)$ (Eq.\,(\ref{Eq:Int_gamma_z}), see the solid curves in Fig.~\ref{fig:Int_gamma_z}) and perform the integral over redshift using standard quadrature integration methods.
\end{enumerate}
 
Finally, we note that the expectation value for the HD correlation itself is given by the HD curve times the squared strain $h^2$ as a normalization factor, see Eq.~\eqref{Eq:HD_strain}. Therefore, in the numerical results presented, all variance contributions are divided through this normalization factor for comparison with the HD curve itself.

\subsection{Results} \label{sec:results_results}

\begin{figure}
    \centering
    \includegraphics[height=6.cm]{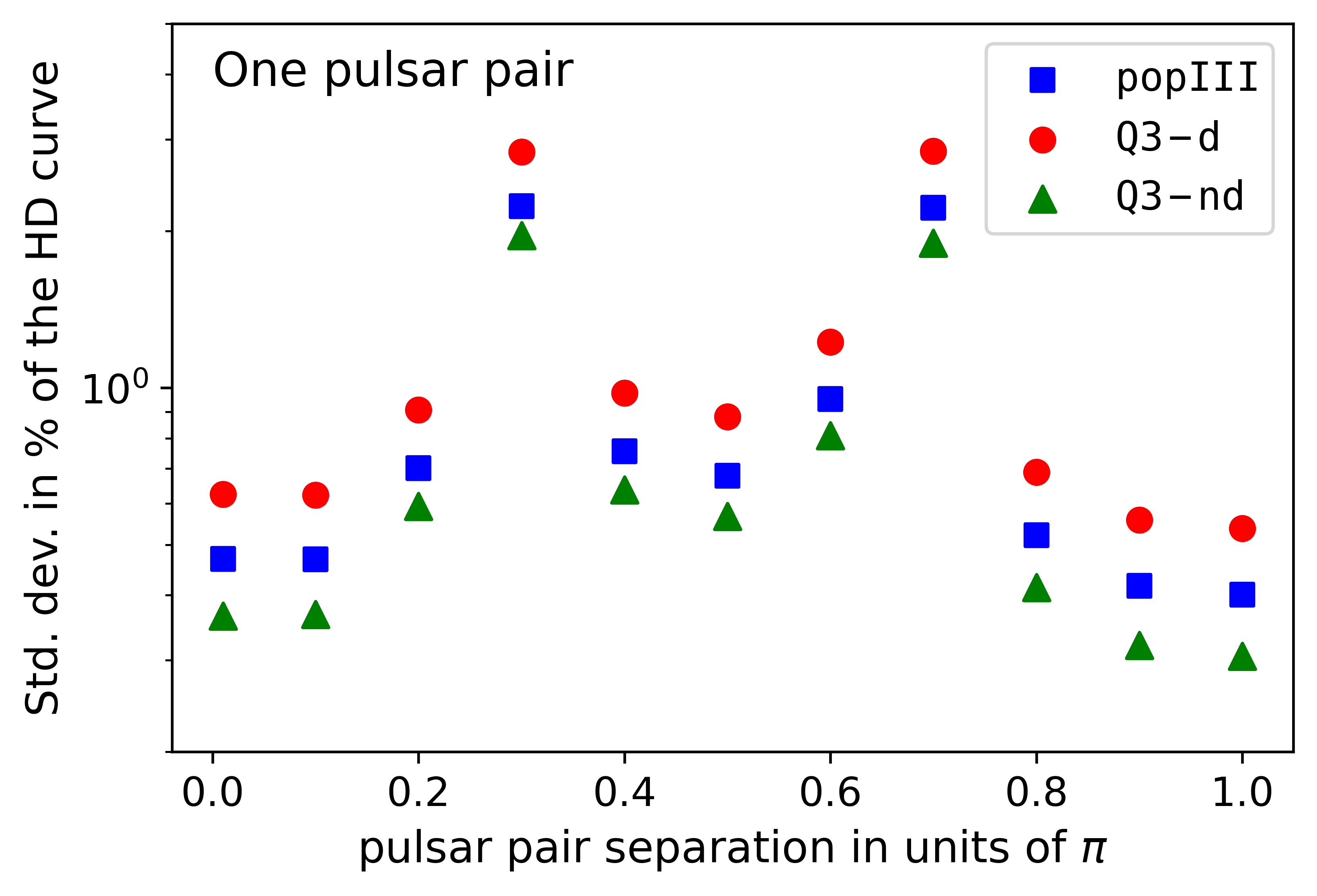}
    \includegraphics[height=6.cm]{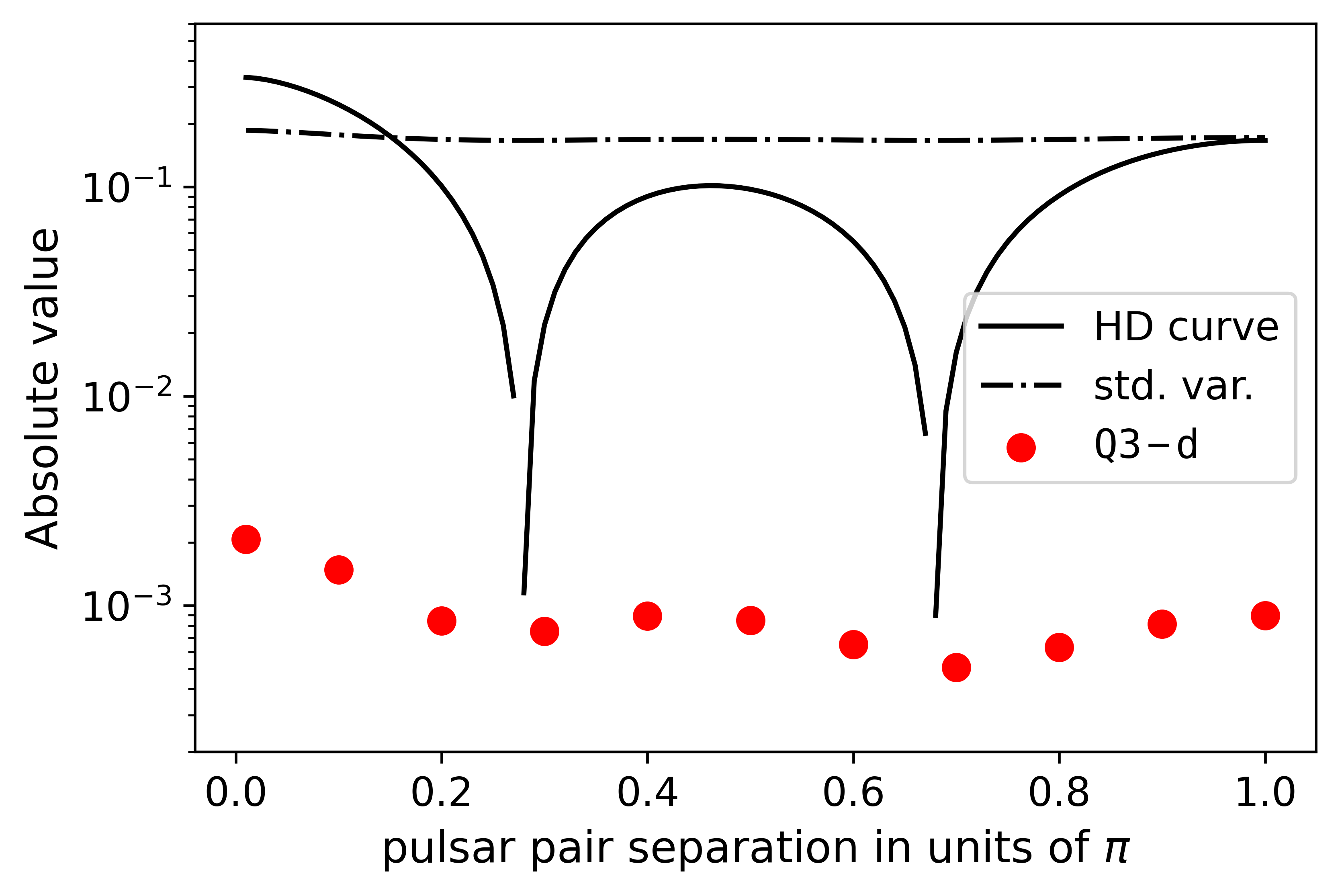}
    \caption{\textit{Left panel:} We show the standard deviation caused by galaxy clustering, divided by the squared strain $h^2$, in percent of the HD curve for the three different astrophysical models in consideration. \textit{Right panel:} We show the absolute value of the HD curve, the size of the standard deviation (square root of the variance) due to the standard variance calculated in Ref.~\cite{Allen:2022dzg}, along with the galaxy clustering contribution (for the {\tt Q3-d} model) calculated in this work.}\label{fig:results_onepair}
\end{figure}

\begin{table*}[htb!!!]
  \centering
  \renewcommand{\arraystretch}{1.2}
		\begin{tabular}{c|| c c c c c c c c c c c}
                Pair separation & 0.01 & 0.1 & 0.2 & 0.3 & 0.4 & 0.5 & 0.6 & 0.7 & 0.8 & 0.9& 1.0 \\
				\hline \hline     
				Standard variance & 56 & 74 & 181 & 629 & 184 & 174 & 314 & 938 & 183 & 116 & 103 \\
				 Clustering variance &  0.63 & 0.62 & 0.91 & 2.84 & 0.97 & 0.88 & 1.23 & 2.85 & 0.69 & 0.56 &  0.54  \\ 
       \hline
				Irr.~standard variance & 7  & 8 & 12 &  9 &  9 &
       12 & 16 & 13 & 12 & 14 & 15  \\
    Irr.~clustering variance & 0.10 & 0.11 & 0.13 & 0.11 & 0.12 &  0.13 & 0.14 & 0.12 & 0.13 & 0.13 & 0.13 \\
    \hline \hline
		\end{tabular}
  \caption{Standard deviation in percent of the HD curve caused by the various contributions summarized in Tab.~\ref{TableBig}. The 11 different pulsar pair separations are given in units of $\pi$. For the clustering variance and pulsar-averaged clustering variance, we assume the {\tt Q3-d} model, leading to slightly higher results than the {\tt popIII} and {\tt Q3-nd} models (see left panels of Figs.~\ref{fig:results_onepair} and~\ref{fig:results_irr}). We only report 11 distinct pulsar pair separations, as the integrals over angles for the clustering variance and pulsar-averaged clustering variance (Eqs.~\eqref{Eq:2ndmom_Limber} and~\eqref{eq:irredclustvariance_Limber}, respectively) are performed numerically.} 
   \label{Table:results}
\end{table*}

In the left panel of Fig.~\ref{fig:results_onepair}, we show the standard deviation caused by galaxy clustering in percent of the HD curve, for the three different astrophysical models in consideration. In the right panel, we compare the different variance contributions to the HD curve itself: for all pulsar pair separations, the standard variance is the dominant contributions, of the order of the HD curve itself, while the clustering variance reaches at most percent-level. These results are shown for the {\tt Q3-d} model, but the same holds for all the three astrophysical models under study. Indeed, we see from the left panel that the differences between the three models are small, at most of the order of $10\%$, with the {\tt Q3-d} model leading to a slightly larger clustering  variance than the other two models. We have also evaluated that the term proportional to $G(z)^2$ in Eq.~\eqref{Eq:2ndmom_Limber} is about 4 times larger than the term proportional to $\Gamma(z,z)$. Since the term proportional to $G(z)^2$ is independent of observational time $T$, stochasticity remains even in the limit of infinite observational time due to galaxy clustering, i.e.~due to the fact that the GW intensity is not uniform on the sky.     

\begin{figure}
    \centering
    \includegraphics[height=5.8cm]{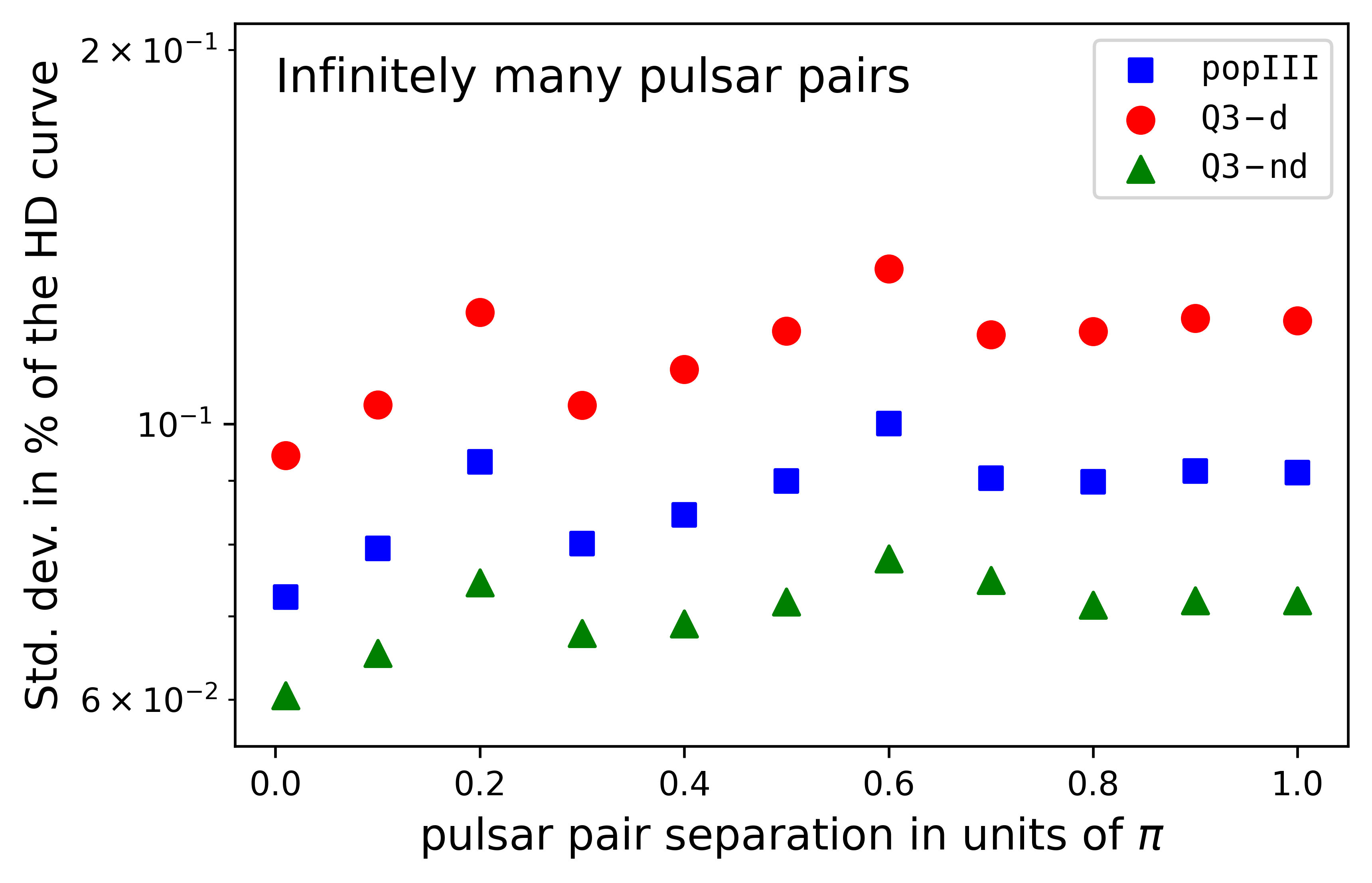}
    \includegraphics[height=5.8cm]{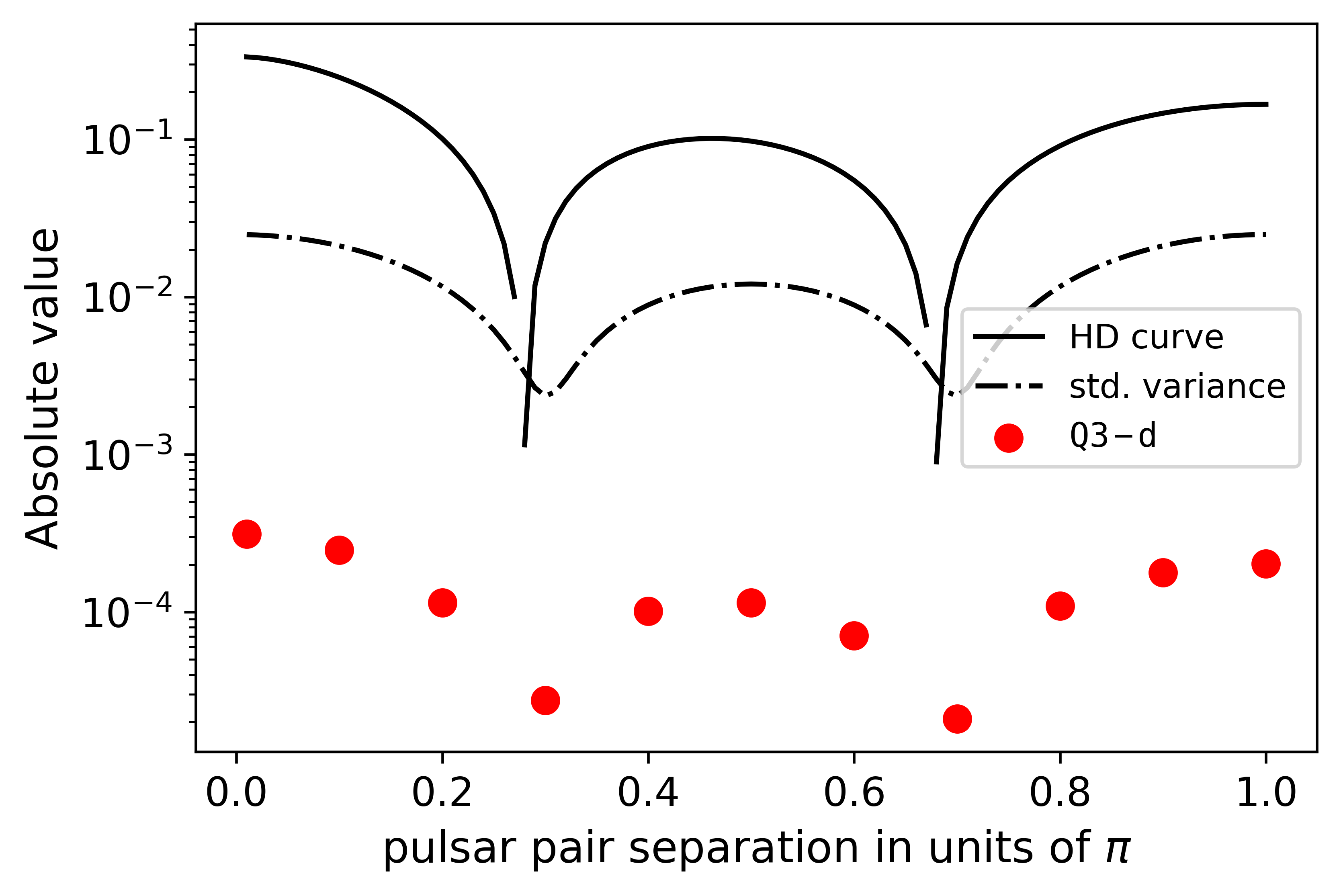}
    \caption{Same as Fig.~\ref{fig:results_onepair}, but for the case of infinitely many pulsar pairs instead of on pair;  \textit{Left panel:} We show the \textit{pulsar-averaged} standard deviation caused by galaxy clustering, divided by the squared strain $h^2$, in percent of the HD curve for the three different astrophysical models in consideration. \textit{Right panel:} We show the absolute value of the HD curve, the size of the standard deviation (square root of the variance) due to the \textit{pulsar-averaged} standard variance calculated in Ref.~\cite{Allen:2022dzg}, along with the \textit{pulsar-averaged} galaxy clustering contribution (for the {\tt Q3-d} model) calculated in this work.}\label{fig:results_irr}
\end{figure}

In Appendix~\ref{App:Shot_pulsar_variance}, we show in detail the calculation of the pulsar-averaged galaxy clustering variance (obtained by averaging over an infinite number of pulsar pairs, at fixed angular separation). Here, we present the numerical results for this case in Fig.~\ref{fig:results_irr}. As can be seen in Table~\ref{Table:results}, we find a suppression by factors $4-26$ for the clustering variance, depending on the pulsar separation (and assuming the {\tt Q3-d} model, although results are similar for all models). Hence, performing the pulsar pair averaging largely counterbalances the stochasticity arising from the anisotropy in GW intensity. More specifically, the HD curve would be exactly recovered in the limit of infinitely many pulsar pairs and an infinitely large observational time. The remaining term $\propto \Gamma(z,z)$, which depends on observational time, is as well altered comparing the single pulsar pair and pulsar-averaged case. In particular, the dependence on the pulsar pair separation is not the same, leading to the variation in suppression factors. In practice, the variance of the HD correlation in a realistic PTA experiment (with a finite number of pulsar pairs for each angular separation) will lie in between the results of Fig.~\ref{fig:results_onepair} and the pulsar-averaged variance results in Fig.~\ref{fig:results_irr}. 
    
We stress that while the equations that we derived are generic, some astrophysical simplifications have been made in our analysis. More specifically, we have assumed no galaxy bias, and that galaxy mergers directly trace mergers of central black holes. Hence, SMBH mergers are treated as perfect tracers of the underlying matter field in our computation. However, as discussed in Section~\ref{sec:bias}, we do not expect an increase of the clustering variance to a level comparable to the standard variance (which is roughly 2 orders of magnitude larger than the clustering variance, see Table~\ref{Table:results}) by lifting these assumptions. {Likewise, while we did not include a frequency sensitivity function, it is unlikely that such detector specifications would substantially change the order of magnitude of the clustering variance compared to the standard variance.} Moreover, even though we applied a standard procedure for large-scale structure observables, small errors could arise e.g.~arise due the fact that the Limber approximation does not precisely describe clustering on very large angular scales. In the case of galaxy clustering, the Limber approximation has been well-studied and is known to work particularly well for broad bins at low redshift, reaching percent-level precision at angular scales $l\gtrsim 10$ (see e.g.\ Fig.~2 of Ref.~\cite{LoVerde:2008re}). Even at the largest scales, $l<10$, the difference between the Limber approximated and precise results is at most at the level of $10\%$. Thus, as the Limber approximation has a high precision for large-scale structure observables, we do not expect it to have a large impact on our results. Another imprecision arises as we apply a sparse sampling in $z$ of the functions $\text{Int}(\gamma,z)$. Since they show a smooth behavior on a double-logarithmic scale, the resulting numerical errors are expected to be small. To conclude, given that the galaxy clustering variance remains significantly below the standard variance, all simplifications made in our analysis are well justified.   

\subsection{Comparison to Allen et al.~(2024)}

\begin{figure}
    \centering
    \includegraphics[width=.495\textwidth]{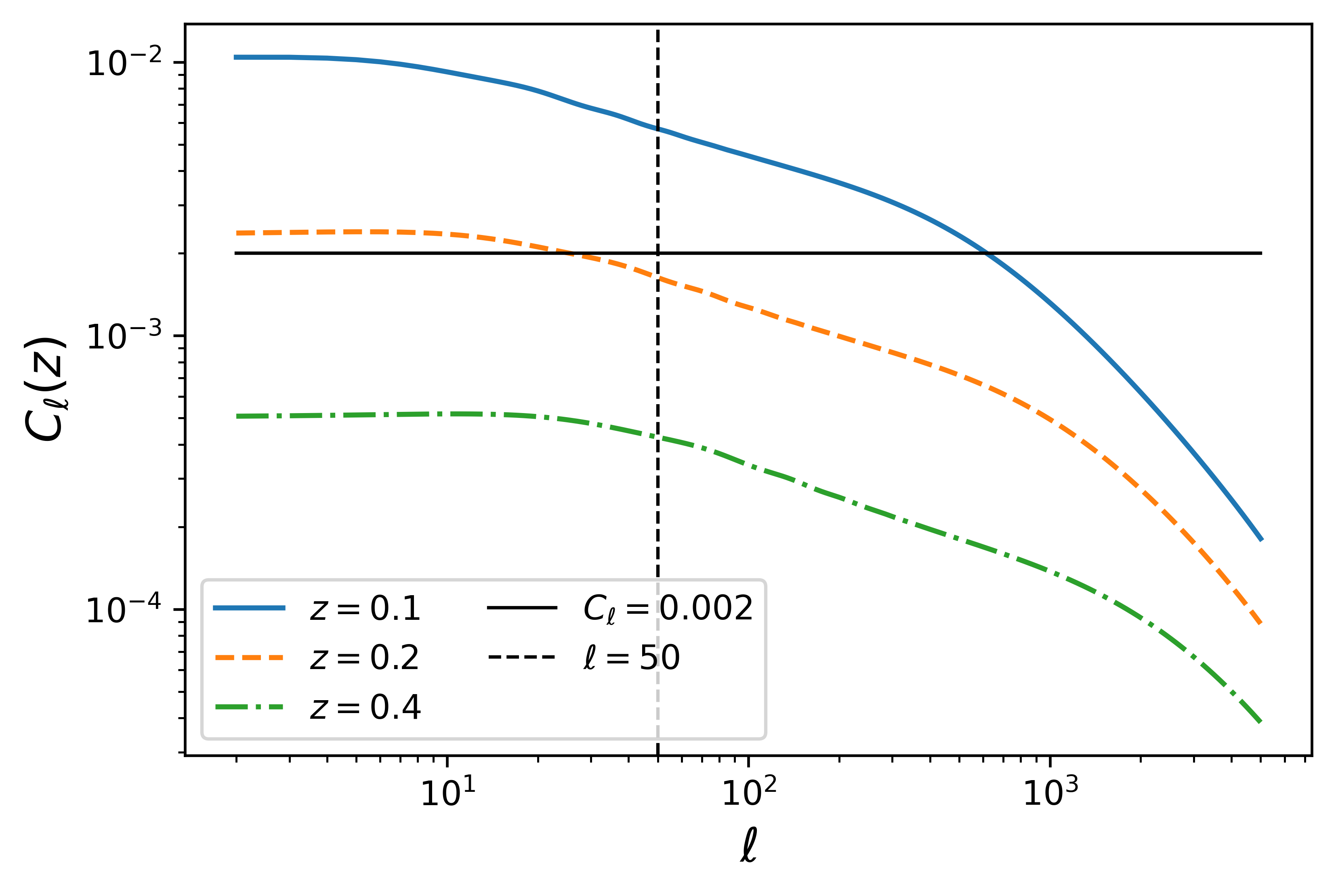}
    \includegraphics[width=.495\textwidth]{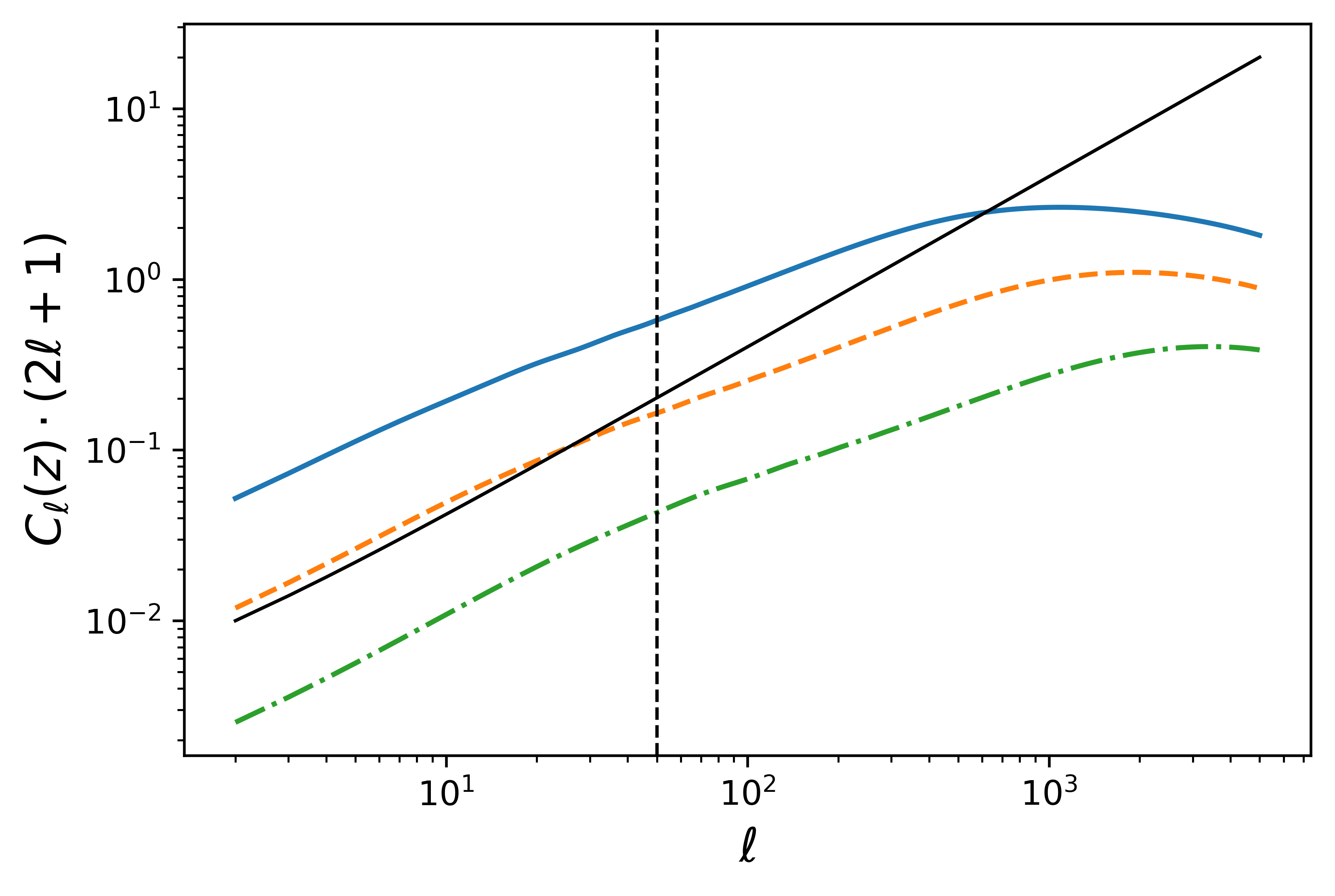}
    \caption{\textit{Left panel:} Angular power spectrum $C_\ell(z=z')$ of matter density correlations at various values of $z$. The constant value $C_\ell=0.002$ and the cut-off of $\ell_{\rm max}=50$ suggested in Ref.~\cite{Allen:2024mtn} is shown as well. \textit{Right panel:} Same as left panel, but with the angular power spectra multiplied by the factor $(2\ell+1)$.}\label{fig:Cl}
\end{figure}

A recent work, Ref.~\cite{Allen:2024mtn}, has as well studied the impact of galaxy clustering on the HD correlation, applying a spherical harmonic analysis. In particular, they choose to work with the angular power spectrum $C_\ell$ of the matter correlations (see Eq.~\eqref{Eq:Cell}). An expression for an upper limit of the clustering contribution in the pulsar-averaged case has been derived under the crude assumption that $C_\ell$ is constant for all $\ell$. Then, setting this constant value to $C_\ell=0.002$, an upper bound of one percent has been found. Employing a simple model where this constant value is cut off and set to zero above $\ell_{\rm max}=50$, the authors argue that the galaxy clustering contribution needs to stay below this percent level. In the left panel of Fig.~\ref{fig:Cl}, we plot the $C_l$ obtained from \texttt{CLASS} at redshifts $z=0.1,0.2,0.4$ (setting $z'=z$ in Eq.~\eqref{Eq:Cell}). The horizontal line corresponding to a constant $C_\ell=0.002$ is as well included, along with the dashed vertical line corresponding to a cut-off at $\ell=50$. We see that, for the lower two redshifts, the $C_\ell$ are above the horizontal line for low angular multipoles. Let us recall that, due to the steep decline of the integrals $\text{Int}(\gamma,z)$ with $z$ and the fact that the astrophysical kernel $b_{\rm GW}(z)$ peaks at $z\sim 0.1$, about $95\%$ of the contribution to the clustering variance arises from redshifts $z\leq 0.1$. The dependence on redshift is not evident in Ref.~\cite{Allen:2024mtn}, while in principle, the clustering contribution should be obtained as the integral over $z$ and $z'$ (see Eq.~(36) in Ref.~\cite{Grimm:2024lfj}), taking the redshift dependence of both $C_\ell(z,z')$ and $b_{\rm GW}(z)b_{\rm GW}(z')$ into account. Therefore, a constant $C_\ell=0.002$ does not properly capture the redshift-dependence and the angular dependence of the angular power spectrum.

We also note that the angular power spectrum $C_l$ cannot be cut at any low multipole. Indeed, while the $C_l$ appear to be strictly decreasing with higher $l$, the strength of galaxy correlation is affected by the product $(2l+1)C_l$, as can be seen from Eq.~\eqref{Eq:Cell}. This product, shown in the right panel of Fig.~\ref{fig:Cl}, increases up to $l\approx 10^3$, reflecting the fact that correlations are stronger at small angular separation (i.e.\ large $l$). For this reason, a priori we do not expect low multipoles $l\lesssim 50$ to be the major contribution to the clustering contribution. All multipoles should be consistently included, weighted by the angular dependence of the functions $\chi_{ab}(\bn)$ that will determine the impact of different multipoles (see Eq.~(36) in Ref.~\cite{Grimm:2024lfj} or Eq.~(6.2) in Ref.~\cite{Allen:2024mtn}). In our case, we work with the galaxy power spectrum in $k$-space instead of the angular power spectrum, and we take the full shape and redshift-dependence of the matter power spectrum $P_m(z,k)$ into account. With this we can precisely compute the galaxy clustering variance, which we find to be at the level of $0.1\%$ in the pulsar-averaged case, i.e.\ one order of magnitude below the upper bound found in~\cite{Allen:2024mtn}. Our methods are readily extendable to study more practical statistical tools to extract the galaxy clustering information from PTA surveys.  For this future task, the scale-dependence of galaxy clustering will be crucial.

\section{Conclusion} \label{sec:conclusion}

In this work, based on the theoretical formalism established in Ref.~\cite{Grimm:2024lfj}, we  performed numerical calculations for the galaxy clustering contribution to the variance of the HD correlation. For the case of one pulsar pair per angular separation, we have found this contribution to be at the level of $0.5-3\%$, depending on the pulsar pair separation. This is substantially smaller than the standard variance, calculated in the absence of anisotropies, which is of the order of the HD curve itself. 

In our work we considered three different astrophysical models to describe the distribution of SMBH GW sources, and we found  qualitatively similar results for all three  models. Our computation in the main part of this work is done for the case of one pulsar pair for each separation, hence the estimate we provide for the galaxy clustering contribution to the variance are an upper bound to what one would obtain in realistic PTA surveys, which typically contain more than one pulsar pair at a given separation. We have also calculated, in Appendix \ref{App:Shot_pulsar_variance}, the pulsar-averaged galaxy clustering variance, obtained by averaging over an infinite number of pulsar pairs at each angular separation. In this case, providing a lower limit for the variance in a realistic observation, both variance contributions (standard and clustering) are suppressed by roughly an order of magnitude compared to the case of one pulsar pair. As a consequence, the galaxy clustering contribution is small compared to the standard contribution as well in this case. Realistic estimates of the variance for finite pulsar pair numbers will lie in between the single pair and infinite pair case (see Fig.~9 in Ref.~\cite{Allen:2022ksj}), but the exact number of pulsar pairs observed will not change the fact that the clustering contribution to the variance remains subdominant with respect to the standard variance. 

Potential sources of astrophysical and numerical errors are discussed at the end of Section~\ref{sec:results}, although none of them are expected to lift this contribution above the percent level.  Therefore, we can safely conclude that the clustering variance is not a major source of uncertainty for current and future PTA surveys. In fact, even if future surveys make a significant improvement in precision with an increase of pulsar pairs and observational time (see e.g.\ Ref.~\cite{Babak:2024yhu}), the variance on the HD correlation will be dominated by the standard variance (calculated in Ref.~\cite{Allen:2022dzg} for an isotropic distribution of GW sources).

In this work we have treated GW source clustering as a source of contamination. However, it is interesting to rather consider it as a signal: the fact that sources are clustered leads to anisotropies in the observed energy density. Different source populations are expected to lead to different levels of anisotropy, hence anisotropies are an interesting observable to look at to distinguish different scenarios for the origin of the SGWB. On the observational side, mapping techniques are already in place, and forecasts have shown that the ability of PTAs to map a background will significantly increase in the future, as the number of pulsar monitored and the observation time will increase~\cite{Depta:2024ykq}. Predictions for the angular power spectrum for different astrophysical scenarios are under study. Moreover, another interesting avenue lies in the cross-correlation between PTAs and large-scale structure tracers, see Ref.~\cite{Semenzato:2024mtn} for a simulation-based approach. Thus, while galaxy clustering does not appreciably affect the variance of the HD correlation, further investigations on synergies between PTAs and galaxy clustering may lead to promising results.\\

\begin{center}
\textbf{Acknowledgements}
\end{center}

 We thank Ruth Durrer, Gabriele Franciolini, Cyril Pitrou, Alberto Sesana, Gianmassimo Tasinato and Marta Volonteri for valuable discussions. We are grateful to Maria Berti, Enea Di Dio and Francesco Iacovelli for useful advice on numerical methods. We acknowledge
the use of the HPC cluster Yggdrasil at the University of Geneva for conducting our numerical evaluations. N.G. and C.B. acknowledge funding from the European Research Council (ERC) under the European Union’s Horizon 2020 research and innovation program (Grant agreement No.~863929; project title ``Testing the law of gravity with novel large-scale structure observables"). The work of G.C. is supported by CNRS and G.C. and M.P. acknowledge support from the Swiss National Science Foundation (Ambizione grant, ``Gravitational wave propagation in the clustered universe").

\bibliographystyle{Bonvinetal}
\bibliography{PTASpectrumRefs}

\ifx\mcitethebibliography\mciteundefinedmacro
\PackageError{LHCb.bst}{mciteplus.sty has not been loaded}
{This bibstyle requires the use of the mciteplus package.}\fi
\providecommand{\href}[2]{#2}
\begin{mcitethebibliography}{10}
\mciteSetBstSublistMode{n}
\mciteSetBstMaxWidthForm{subitem}{\alph{mcitesubitemcount})}
\mciteSetBstSublistLabelBeginEnd{\mcitemaxwidthsubitemform\space}
{\relax}{\relax}

\bibitem{NANOGrav:2023gor}
NANOGrav, G.~Agazie {\em et~al.}, \ifthenelse{\boolean{articletitles}}{\emph{{The NANOGrav 15 yr Data Set: Evidence for a Gravitational-wave Background}}, }{}\href{https://doi.org/10.3847/2041-8213/acdac6}{Astrophys.\ J.\ Lett.\  \textbf{951} (2023) L8}, \href{http://arxiv.org/abs/2306.16213}{{\normalfont\ttfamily arXiv:2306.16213}}\relax
\mciteBstWouldAddEndPuncttrue
\mciteSetBstMidEndSepPunct{\mcitedefaultmidpunct}
{\mcitedefaultendpunct}{\mcitedefaultseppunct}\relax
\EndOfBibitem
\bibitem{Reardon:2023gzh}
D.~J. Reardon {\em et~al.}, \ifthenelse{\boolean{articletitles}}{\emph{{Search for an Isotropic Gravitational-wave Background with the Parkes Pulsar Timing Array}}, }{}\href{https://doi.org/10.3847/2041-8213/acdd02}{Astrophys.\ J.\ Lett.\  \textbf{951} (2023) L6}, \href{http://arxiv.org/abs/2306.16215}{{\normalfont\ttfamily arXiv:2306.16215}}\relax
\mciteBstWouldAddEndPuncttrue
\mciteSetBstMidEndSepPunct{\mcitedefaultmidpunct}
{\mcitedefaultendpunct}{\mcitedefaultseppunct}\relax
\EndOfBibitem
\bibitem{EPTA:2023sfo}
EPTA, J.~Antoniadis {\em et~al.}, \ifthenelse{\boolean{articletitles}}{\emph{{The second data release from the European Pulsar Timing Array I. The dataset and timing analysis}}, }{}\href{https://doi.org/10.1051/0004-6361/202346841}{Astron.\ Astrophys.\  \textbf{678} (2023) A48}, \href{http://arxiv.org/abs/2306.16224}{{\normalfont\ttfamily arXiv:2306.16224}}\relax
\mciteBstWouldAddEndPuncttrue
\mciteSetBstMidEndSepPunct{\mcitedefaultmidpunct}
{\mcitedefaultendpunct}{\mcitedefaultseppunct}\relax
\EndOfBibitem
\bibitem{Xu:2023wog}
H.~Xu {\em et~al.}, \ifthenelse{\boolean{articletitles}}{\emph{{Searching for the Nano-Hertz Stochastic Gravitational Wave Background with the Chinese Pulsar Timing Array Data Release I}}, }{}\href{https://doi.org/10.1088/1674-4527/acdfa5}{Res.\ Astron.\ Astrophys.\  \textbf{23} (2023) 075024}, \href{http://arxiv.org/abs/2306.16216}{{\normalfont\ttfamily arXiv:2306.16216}}\relax
\mciteBstWouldAddEndPuncttrue
\mciteSetBstMidEndSepPunct{\mcitedefaultmidpunct}
{\mcitedefaultendpunct}{\mcitedefaultseppunct}\relax
\EndOfBibitem
\bibitem{Hellings:1983fr}
R.~w. Hellings and G.~s. Downs, \ifthenelse{\boolean{articletitles}}{\emph{{UPPER LIMITS ON THE ISOTROPIC GRAVITATIONAL RADIATION BACKGROUND FROM PULSAR TIMING ANALYSIS}}, }{}\href{https://doi.org/10.1086/183954}{Astrophys.\ J.\ Lett.\  \textbf{265} (1983) L39}\relax
\mciteBstWouldAddEndPuncttrue
\mciteSetBstMidEndSepPunct{\mcitedefaultmidpunct}
{\mcitedefaultendpunct}{\mcitedefaultseppunct}\relax
\EndOfBibitem
\bibitem{Allen:2022bjz}
B.~Allen, \ifthenelse{\boolean{articletitles}}{\emph{{Will pulsar timing arrays observe the hellings and downs correlation curve?}}, }{}Frascati Phys.\ Ser.\  \textbf{74} (2022) 65\relax
\mciteBstWouldAddEndPuncttrue
\mciteSetBstMidEndSepPunct{\mcitedefaultmidpunct}
{\mcitedefaultendpunct}{\mcitedefaultseppunct}\relax
\EndOfBibitem
\bibitem{Allen:2022dzg}
B.~Allen, \ifthenelse{\boolean{articletitles}}{\emph{{Variance of the Hellings-Downs correlation}}, }{}\href{https://doi.org/10.1103/PhysRevD.107.043018}{Phys.\ Rev.\ D \textbf{107} (2023) 043018}, \href{http://arxiv.org/abs/2205.05637}{{\normalfont\ttfamily arXiv:2205.05637}}\relax
\mciteBstWouldAddEndPuncttrue
\mciteSetBstMidEndSepPunct{\mcitedefaultmidpunct}
{\mcitedefaultendpunct}{\mcitedefaultseppunct}\relax
\EndOfBibitem
\bibitem{Allen:2022ksj}
B.~Allen and J.~D. Romano, \ifthenelse{\boolean{articletitles}}{\emph{{Hellings and Downs correlation of an arbitrary set of pulsars}}, }{}\href{https://doi.org/10.1103/PhysRevD.108.043026}{Phys.\ Rev.\ D \textbf{108} (2023) 043026}, \href{http://arxiv.org/abs/2208.07230}{{\normalfont\ttfamily arXiv:2208.07230}}\relax
\mciteBstWouldAddEndPuncttrue
\mciteSetBstMidEndSepPunct{\mcitedefaultmidpunct}
{\mcitedefaultendpunct}{\mcitedefaultseppunct}\relax
\EndOfBibitem
\bibitem{Romano:2023zhb}
J.~D. Romano and B.~Allen, \ifthenelse{\boolean{articletitles}}{\emph{{Answers to frequently asked questions about the pulsar timing array Hellings and Downs curve}}, }{}\href{https://doi.org/10.1088/1361-6382/ad4c4c}{Class.\ Quant.\ Grav.\  \textbf{41} (2024) 175008}, \href{http://arxiv.org/abs/2308.05847}{{\normalfont\ttfamily arXiv:2308.05847}}\relax
\mciteBstWouldAddEndPuncttrue
\mciteSetBstMidEndSepPunct{\mcitedefaultmidpunct}
{\mcitedefaultendpunct}{\mcitedefaultseppunct}\relax
\EndOfBibitem
\bibitem{Allen:2024rqk}
B.~Allen and S.~Valtolina, \ifthenelse{\boolean{articletitles}}{\emph{{Pulsar timing array source ensembles}}, }{}\href{https://doi.org/10.1103/PhysRevD.109.083038}{Phys.\ Rev.\ D \textbf{109} (2024) 083038}, \href{http://arxiv.org/abs/2401.14329}{{\normalfont\ttfamily arXiv:2401.14329}}\relax
\mciteBstWouldAddEndPuncttrue
\mciteSetBstMidEndSepPunct{\mcitedefaultmidpunct}
{\mcitedefaultendpunct}{\mcitedefaultseppunct}\relax
\EndOfBibitem
\bibitem{Bernardo:2022xzl}
R.~C. Bernardo and K.-W. Ng, \ifthenelse{\boolean{articletitles}}{\emph{{Pulsar and cosmic variances of pulsar timing-array correlation measurements of the stochastic gravitational wave background}}, }{}\href{https://doi.org/10.1088/1475-7516/2022/11/046}{JCAP \textbf{11} (2022) 046}, \href{http://arxiv.org/abs/2209.14834}{{\normalfont\ttfamily arXiv:2209.14834}}\relax
\mciteBstWouldAddEndPuncttrue
\mciteSetBstMidEndSepPunct{\mcitedefaultmidpunct}
{\mcitedefaultendpunct}{\mcitedefaultseppunct}\relax
\EndOfBibitem
\bibitem{Grimm:2024lfj}
N.~Grimm, M.~Pijnenburg, G.~Cusin, and C.~Bonvin, \ifthenelse{\boolean{articletitles}}{\emph{{The impact of large-scale galaxy clustering on the variance of the Hellings-Downs correlation: theoretical framework}}, }{}\href{https://doi.org/10.1088/1475-7516/2025/03/011}{JCAP \textbf{03} (2025) 011}, \href{http://arxiv.org/abs/2404.05670}{{\normalfont\ttfamily arXiv:2404.05670}}\relax
\mciteBstWouldAddEndPuncttrue
\mciteSetBstMidEndSepPunct{\mcitedefaultmidpunct}
{\mcitedefaultendpunct}{\mcitedefaultseppunct}\relax
\EndOfBibitem
\bibitem{Allen:1996gp}
B.~Allen and A.~C. Ottewill, \ifthenelse{\boolean{articletitles}}{\emph{{Detection of anisotropies in the gravitational wave stochastic background}}, }{}\href{https://doi.org/10.1103/PhysRevD.56.545}{Phys.\ Rev.\ D \textbf{56} (1997) 545}, \href{http://arxiv.org/abs/gr-qc/9607068}{{\normalfont\ttfamily arXiv:gr-qc/9607068}}\relax
\mciteBstWouldAddEndPuncttrue
\mciteSetBstMidEndSepPunct{\mcitedefaultmidpunct}
{\mcitedefaultendpunct}{\mcitedefaultseppunct}\relax
\EndOfBibitem
\bibitem{Cusin:2017fwz}
G.~Cusin, C.~Pitrou, and J.-P. Uzan, \ifthenelse{\boolean{articletitles}}{\emph{{Anisotropy of the astrophysical gravitational wave background: Analytic expression of the angular power spectrum and correlation with cosmological observations}}, }{}\href{https://doi.org/10.1103/PhysRevD.96.103019}{Phys.\ Rev.\  \textbf{D96} (2017) 103019}, \href{http://arxiv.org/abs/1704.06184}{{\normalfont\ttfamily arXiv:1704.06184}}\relax
\mciteBstWouldAddEndPuncttrue
\mciteSetBstMidEndSepPunct{\mcitedefaultmidpunct}
{\mcitedefaultendpunct}{\mcitedefaultseppunct}\relax
\EndOfBibitem
\bibitem{Cusin:2017mjm}
G.~Cusin, C.~Pitrou, and J.-P. Uzan, \ifthenelse{\boolean{articletitles}}{\emph{{The signal of the gravitational wave background and the angular correlation of its energy density}}, }{}\href{https://doi.org/10.1103/PhysRevD.97.123527}{Phys.\ Rev.\  \textbf{D97} (2018) 123527}, \href{http://arxiv.org/abs/1711.11345}{{\normalfont\ttfamily arXiv:1711.11345}}\relax
\mciteBstWouldAddEndPuncttrue
\mciteSetBstMidEndSepPunct{\mcitedefaultmidpunct}
{\mcitedefaultendpunct}{\mcitedefaultseppunct}\relax
\EndOfBibitem
\bibitem{Cusin:2018rsq}
G.~Cusin, I.~Dvorkin, C.~Pitrou, and J.-P. Uzan, \ifthenelse{\boolean{articletitles}}{\emph{{First predictions of the angular power spectrum of the astrophysical gravitational wave background}}, }{}\href{https://doi.org/10.1103/PhysRevLett.120.231101}{Phys.\ Rev.\ Lett.\  \textbf{120} (2018) 231101}, \href{http://arxiv.org/abs/1803.03236}{{\normalfont\ttfamily arXiv:1803.03236}}\relax
\mciteBstWouldAddEndPuncttrue
\mciteSetBstMidEndSepPunct{\mcitedefaultmidpunct}
{\mcitedefaultendpunct}{\mcitedefaultseppunct}\relax
\EndOfBibitem
\bibitem{Cusin:2018avf}
G.~Cusin, R.~Durrer, and P.~G. Ferreira, \ifthenelse{\boolean{articletitles}}{\emph{{Polarization of a stochastic gravitational wave background through diffusion by massive structures}}, }{}\href{https://doi.org/10.1103/PhysRevD.99.023534}{Phys.\ Rev.\ D \textbf{99} (2019) 023534}, \href{http://arxiv.org/abs/1807.10620}{{\normalfont\ttfamily arXiv:1807.10620}}\relax
\mciteBstWouldAddEndPuncttrue
\mciteSetBstMidEndSepPunct{\mcitedefaultmidpunct}
{\mcitedefaultendpunct}{\mcitedefaultseppunct}\relax
\EndOfBibitem
\bibitem{Cusin:2019jpv}
G.~Cusin, I.~Dvorkin, C.~Pitrou, and J.-P. Uzan, \ifthenelse{\boolean{articletitles}}{\emph{{Properties of the stochastic astrophysical gravitational wave background: astrophysical sources dependencies}}, }{}\href{https://doi.org/10.1103/PhysRevD.100.063004}{Phys.\ Rev.\ D \textbf{100} (2019) 063004}, \href{http://arxiv.org/abs/1904.07797}{{\normalfont\ttfamily arXiv:1904.07797}}\relax
\mciteBstWouldAddEndPuncttrue
\mciteSetBstMidEndSepPunct{\mcitedefaultmidpunct}
{\mcitedefaultendpunct}{\mcitedefaultseppunct}\relax
\EndOfBibitem
\bibitem{Cusin:2019jhg}
G.~Cusin, I.~Dvorkin, C.~Pitrou, and J.-P. Uzan, \ifthenelse{\boolean{articletitles}}{\emph{{Stochastic gravitational wave background anisotropies in the mHz band: astrophysical dependencies}}, }{}\href{https://doi.org/10.1093/mnrasl/slz182}{Mon.\ Not.\ Roy.\ Astron.\ Soc.\  \textbf{493} (2020) L1}, \href{http://arxiv.org/abs/1904.07757}{{\normalfont\ttfamily arXiv:1904.07757}}\relax
\mciteBstWouldAddEndPuncttrue
\mciteSetBstMidEndSepPunct{\mcitedefaultmidpunct}
{\mcitedefaultendpunct}{\mcitedefaultseppunct}\relax
\EndOfBibitem
\bibitem{Pitrou:2019rjz}
C.~Pitrou, G.~Cusin, and J.-P. Uzan, \ifthenelse{\boolean{articletitles}}{\emph{{Unified view of anisotropies in the astrophysical gravitational-wave background}}, }{}\href{https://doi.org/10.1103/PhysRevD.101.081301}{Phys.\ Rev.\ D \textbf{101} (2020) 081301}, \href{http://arxiv.org/abs/1910.04645}{{\normalfont\ttfamily arXiv:1910.04645}}\relax
\mciteBstWouldAddEndPuncttrue
\mciteSetBstMidEndSepPunct{\mcitedefaultmidpunct}
{\mcitedefaultendpunct}{\mcitedefaultseppunct}\relax
\EndOfBibitem
\bibitem{Jenkins:2019nks}
A.~C. Jenkins, J.~D. Romano, and M.~Sakellariadou, \ifthenelse{\boolean{articletitles}}{\emph{Estimating the angular power spectrum of the gravitational-wave background in the presence of shot noise}, }{}\href{https://doi.org/10.1103/PhysRevD.100.083501}{Phys.\ Rev.\ D \textbf{100} (2019) 083501}\relax
\mciteBstWouldAddEndPuncttrue
\mciteSetBstMidEndSepPunct{\mcitedefaultmidpunct}
{\mcitedefaultendpunct}{\mcitedefaultseppunct}\relax
\EndOfBibitem
\bibitem{Alonso:2020mva}
D.~Alonso, G.~Cusin, P.~G. Ferreira, and C.~Pitrou, \ifthenelse{\boolean{articletitles}}{\emph{{Detecting the anisotropic astrophysical gravitational wave background in the presence of shot noise through cross-correlations}}, }{}\href{https://doi.org/10.1103/PhysRevD.102.023002}{Phys.\ Rev.\ D \textbf{102} (2020) 023002}, \href{http://arxiv.org/abs/2002.02888}{{\normalfont\ttfamily arXiv:2002.02888}}\relax
\mciteBstWouldAddEndPuncttrue
\mciteSetBstMidEndSepPunct{\mcitedefaultmidpunct}
{\mcitedefaultendpunct}{\mcitedefaultseppunct}\relax
\EndOfBibitem
\bibitem{renzini2022}
A.~I. Renzini, J.~D. Romano, C.~R. Contaldi, and N.~J. Cornish, \ifthenelse{\boolean{articletitles}}{\emph{Comparison of maximum-likelihood mapping methods for gravitational-wave backgrounds}, }{}\href{https://doi.org/10.1103/PhysRevD.105.023519}{Phys.\ Rev.\ D \textbf{105} (2022) 023519}\relax
\mciteBstWouldAddEndPuncttrue
\mciteSetBstMidEndSepPunct{\mcitedefaultmidpunct}
{\mcitedefaultendpunct}{\mcitedefaultseppunct}\relax
\EndOfBibitem
\bibitem{Mingarelli:2013dsa}
C.~M.~F. Mingarelli, T.~Sidery, I.~Mandel, and A.~Vecchio, \ifthenelse{\boolean{articletitles}}{\emph{{Characterizing gravitational wave stochastic background anisotropy with pulsar timing arrays}}, }{}\href{https://doi.org/10.1103/PhysRevD.88.062005}{Phys.\ Rev.\ D \textbf{88} (2013) 062005}, \href{http://arxiv.org/abs/1306.5394}{{\normalfont\ttfamily arXiv:1306.5394}}\relax
\mciteBstWouldAddEndPuncttrue
\mciteSetBstMidEndSepPunct{\mcitedefaultmidpunct}
{\mcitedefaultendpunct}{\mcitedefaultseppunct}\relax
\EndOfBibitem
\bibitem{Taylor:2013esa}
S.~R. Taylor and J.~R. Gair, \ifthenelse{\boolean{articletitles}}{\emph{{Searching For Anisotropic Gravitational-wave Backgrounds Using Pulsar Timing Arrays}}, }{}\href{https://doi.org/10.1103/PhysRevD.88.084001}{Phys.\ Rev.\ D \textbf{88} (2013) 084001}, \href{http://arxiv.org/abs/1306.5395}{{\normalfont\ttfamily arXiv:1306.5395}}\relax
\mciteBstWouldAddEndPuncttrue
\mciteSetBstMidEndSepPunct{\mcitedefaultmidpunct}
{\mcitedefaultendpunct}{\mcitedefaultseppunct}\relax
\EndOfBibitem
\bibitem{Gair:2014rwa}
J.~Gair, J.~D. Romano, S.~Taylor, and C.~M.~F. Mingarelli, \ifthenelse{\boolean{articletitles}}{\emph{{Mapping gravitational-wave backgrounds using methods from CMB analysis: Application to pulsar timing arrays}}, }{}\href{https://doi.org/10.1103/PhysRevD.90.082001}{Phys.\ Rev.\  \textbf{D90} (2014) 082001}, \href{http://arxiv.org/abs/1406.4664}{{\normalfont\ttfamily arXiv:1406.4664}}\relax
\mciteBstWouldAddEndPuncttrue
\mciteSetBstMidEndSepPunct{\mcitedefaultmidpunct}
{\mcitedefaultendpunct}{\mcitedefaultseppunct}\relax
\EndOfBibitem
\bibitem{Ali-Haimoud:2016mbv}
Y.~Ali-Haïmoud and M.~Kamionkowski, \ifthenelse{\boolean{articletitles}}{\emph{{Cosmic microwave background limits on accreting primordial black holes}}, }{}\href{https://doi.org/10.1103/PhysRevD.95.043534}{Phys.\ Rev.\  \textbf{D95} (2017) 043534}, \href{http://arxiv.org/abs/1612.05644}{{\normalfont\ttfamily arXiv:1612.05644}}\relax
\mciteBstWouldAddEndPuncttrue
\mciteSetBstMidEndSepPunct{\mcitedefaultmidpunct}
{\mcitedefaultendpunct}{\mcitedefaultseppunct}\relax
\EndOfBibitem
\bibitem{Cruz:2024svc}
N.~M.~J. Cruz, A.~Malhotra, G.~Tasinato, and I.~Zavala, \ifthenelse{\boolean{articletitles}}{\emph{{Measuring kinematic anisotropies with pulsar timing arrays}}, }{}\href{http://arxiv.org/abs/2402.17312}{{\normalfont\ttfamily arXiv:2402.17312}}\relax
\mciteBstWouldAddEndPuncttrue
\mciteSetBstMidEndSepPunct{\mcitedefaultmidpunct}
{\mcitedefaultendpunct}{\mcitedefaultseppunct}\relax
\EndOfBibitem
\bibitem{BarausseCatalogue}
E.~Barausse, \ifthenelse{\boolean{articletitles}}{\emph{The evolution of massive black holes and their spins in their galactic hosts}, }{}Monthly Notices of the Royal Astronomical Society \textbf{423} (2012) 2533\relax
\mciteBstWouldAddEndPuncttrue
\mciteSetBstMidEndSepPunct{\mcitedefaultmidpunct}
{\mcitedefaultendpunct}{\mcitedefaultseppunct}\relax
\EndOfBibitem
\bibitem{Klein:2015hvg}
A.~Klein {\em et~al.}, \ifthenelse{\boolean{articletitles}}{\emph{{Science with the space-based interferometer eLISA: Supermassive black hole binaries}}, }{}\href{https://doi.org/10.1103/PhysRevD.93.024003}{Phys.\ Rev.\ D \textbf{93} (2016) 024003}, \href{http://arxiv.org/abs/1511.05581}{{\normalfont\ttfamily arXiv:1511.05581}}\relax
\mciteBstWouldAddEndPuncttrue
\mciteSetBstMidEndSepPunct{\mcitedefaultmidpunct}
{\mcitedefaultendpunct}{\mcitedefaultseppunct}\relax
\EndOfBibitem
\bibitem{Pitrou:2024scp}
C.~Pitrou and G.~Cusin, \ifthenelse{\boolean{articletitles}}{\emph{{Mitigating cosmic variance in the Hellings-Downs curve: A cosmic microwave background analogy}}, }{}\href{https://doi.org/10.1103/PhysRevD.111.083544}{Phys.\ Rev.\ D \textbf{111} (2025) 083544}, \href{http://arxiv.org/abs/2412.12073}{{\normalfont\ttfamily arXiv:2412.12073}}\relax
\mciteBstWouldAddEndPuncttrue
\mciteSetBstMidEndSepPunct{\mcitedefaultmidpunct}
{\mcitedefaultendpunct}{\mcitedefaultseppunct}\relax
\EndOfBibitem
\bibitem{Allen:2024uqs}
B.~Allen and J.~D. Romano, \ifthenelse{\boolean{articletitles}}{\emph{{Optimal Reconstruction of the Hellings and Downs Correlation}}, }{}\href{https://doi.org/10.1103/PhysRevLett.134.031401}{Phys.\ Rev.\ Lett.\  \textbf{134} (2025) 031401}, \href{http://arxiv.org/abs/2407.10968}{{\normalfont\ttfamily arXiv:2407.10968}}\relax
\mciteBstWouldAddEndPuncttrue
\mciteSetBstMidEndSepPunct{\mcitedefaultmidpunct}
{\mcitedefaultendpunct}{\mcitedefaultseppunct}\relax
\EndOfBibitem
\bibitem{Allen:2024mtn}
B.~Allen, D.~Agarwal, J.~D. Romano, and S.~Valtolina, \ifthenelse{\boolean{articletitles}}{\emph{{Source anisotropies and pulsar timing arrays}}, }{}\href{https://doi.org/10.1103/PhysRevD.110.123507}{Phys.\ Rev.\ D \textbf{110} (2024) 123507}, \href{http://arxiv.org/abs/2406.16031}{{\normalfont\ttfamily arXiv:2406.16031}}\relax
\mciteBstWouldAddEndPuncttrue
\mciteSetBstMidEndSepPunct{\mcitedefaultmidpunct}
{\mcitedefaultendpunct}{\mcitedefaultseppunct}\relax
\EndOfBibitem
\bibitem{Allen:2024bnk}
B.~Allen, \ifthenelse{\boolean{articletitles}}{\emph{{Pulsar timing array harmonic analysis and source angular correlations}}, }{}\href{https://doi.org/10.1103/PhysRevD.110.043043}{Phys.\ Rev.\ D \textbf{110} (2024) 043043}, \href{http://arxiv.org/abs/2404.05677}{{\normalfont\ttfamily arXiv:2404.05677}}\relax
\mciteBstWouldAddEndPuncttrue
\mciteSetBstMidEndSepPunct{\mcitedefaultmidpunct}
{\mcitedefaultendpunct}{\mcitedefaultseppunct}\relax
\EndOfBibitem
\bibitem{Agarwal:2024hlj}
D.~Agarwal and J.~D. Romano, \ifthenelse{\boolean{articletitles}}{\emph{{Cosmic variance of the Hellings and Downs correlation for ensembles of universes having nonzero angular power spectra}}, }{}\href{https://doi.org/10.1103/PhysRevD.110.043044}{Phys.\ Rev.\ D \textbf{110} (2024) 043044}, \href{http://arxiv.org/abs/2404.08574}{{\normalfont\ttfamily arXiv:2404.08574}}\relax
\mciteBstWouldAddEndPuncttrue
\mciteSetBstMidEndSepPunct{\mcitedefaultmidpunct}
{\mcitedefaultendpunct}{\mcitedefaultseppunct}\relax
\EndOfBibitem
\bibitem{Maggiore:1900zz}
M.~Maggiore, {\em {Gravitational Waves. Vol. 1: Theory and Experiments}}, Oxford Master Series in Physics, Oxford University Press, 2007\relax
\mciteBstWouldAddEndPuncttrue
\mciteSetBstMidEndSepPunct{\mcitedefaultmidpunct}
{\mcitedefaultendpunct}{\mcitedefaultseppunct}\relax
\EndOfBibitem
\bibitem{Dvorkin:2016okx}
I.~Dvorkin, J.-P. Uzan, E.~Vangioni, and J.~Silk, \ifthenelse{\boolean{articletitles}}{\emph{{Synthetic model of the gravitational wave background from evolving binary compact objects}}, }{}\href{https://doi.org/10.1103/PhysRevD.94.103011}{Phys.\ Rev.\  \textbf{D94} (2016) 103011}, \href{http://arxiv.org/abs/1607.06818}{{\normalfont\ttfamily arXiv:1607.06818}}\relax
\mciteBstWouldAddEndPuncttrue
\mciteSetBstMidEndSepPunct{\mcitedefaultmidpunct}
{\mcitedefaultendpunct}{\mcitedefaultseppunct}\relax
\EndOfBibitem
\bibitem{Phinney}
E.~S. Phinney, \ifthenelse{\boolean{articletitles}}{\emph{{A Practical theorem on gravitational wave backgrounds}}, }{}\href{http://arxiv.org/abs/astro-ph/0108028}{{\normalfont\ttfamily arXiv:astro-ph/0108028}}\relax
\mciteBstWouldAddEndPuncttrue
\mciteSetBstMidEndSepPunct{\mcitedefaultmidpunct}
{\mcitedefaultendpunct}{\mcitedefaultseppunct}\relax
\EndOfBibitem
\bibitem{Bonvin:2011bg}
C.~Bonvin and R.~Durrer, \ifthenelse{\boolean{articletitles}}{\emph{{What galaxy surveys really measure}}, }{}\href{https://doi.org/10.1103/PhysRevD.84.063505}{Phys.\ Rev.\  \textbf{D84} (2011) 063505}, \href{http://arxiv.org/abs/1105.5280}{{\normalfont\ttfamily arXiv:1105.5280}}\relax
\mciteBstWouldAddEndPuncttrue
\mciteSetBstMidEndSepPunct{\mcitedefaultmidpunct}
{\mcitedefaultendpunct}{\mcitedefaultseppunct}\relax
\EndOfBibitem
\bibitem{Maggiore:2018sht}
M.~Maggiore, {\em {Gravitational Waves. Vol. 2: Astrophysics and Cosmology}}, Oxford University Press, 2018\relax
\mciteBstWouldAddEndPuncttrue
\mciteSetBstMidEndSepPunct{\mcitedefaultmidpunct}
{\mcitedefaultendpunct}{\mcitedefaultseppunct}\relax
\EndOfBibitem
\bibitem{Sesana:2008mz}
A.~Sesana, A.~Vecchio, and C.~N. Colacino, \ifthenelse{\boolean{articletitles}}{\emph{{The stochastic gravitational-wave background from massive black hole binary systems: implications for observations with Pulsar Timing Arrays}}, }{}\href{https://doi.org/10.1111/j.1365-2966.2008.13682.x}{Mon.\ Not.\ Roy.\ Astron.\ Soc.\  \textbf{390} (2008) 192}, \href{http://arxiv.org/abs/0804.4476}{{\normalfont\ttfamily arXiv:0804.4476}}\relax
\mciteBstWouldAddEndPuncttrue
\mciteSetBstMidEndSepPunct{\mcitedefaultmidpunct}
{\mcitedefaultendpunct}{\mcitedefaultseppunct}\relax
\EndOfBibitem
\bibitem{Sargant_historicalWeighting}
W.~L.~W. {Sargent}, P.~J. {Young}, A.~{Boksenberg}, K.~{Shortridge}, C.~R. {Lynds} {\em et~al.}, \ifthenelse{\boolean{articletitles}}{\emph{{Dynamical evidence for a central mass concentration in the galaxy M87.}}, }{}\href{https://doi.org/10.1086/156077}{\apj \textbf{221} (1978) 731}\relax
\mciteBstWouldAddEndPuncttrue
\mciteSetBstMidEndSepPunct{\mcitedefaultmidpunct}
{\mcitedefaultendpunct}{\mcitedefaultseppunct}\relax
\EndOfBibitem
\bibitem{Blanford_ReverberationMapping}
R.~D. {Blandford} and C.~F. {McKee}, \ifthenelse{\boolean{articletitles}}{\emph{{Reverberation mapping of the emission line regions of Seyfert galaxies and quasars.}}, }{}\href{https://doi.org/10.1086/159843}{\apj \textbf{255} (1982) 419}\relax
\mciteBstWouldAddEndPuncttrue
\mciteSetBstMidEndSepPunct{\mcitedefaultmidpunct}
{\mcitedefaultendpunct}{\mcitedefaultseppunct}\relax
\EndOfBibitem
\bibitem{Peterson_ReverberationMapping}
B.~M. {Peterson}, \ifthenelse{\boolean{articletitles}}{\emph{{Reverberation Mapping of Active Galactic Nuclei}}, }{}\href{https://doi.org/10.1086/133140}{\pasp \textbf{105} (1993) 247}\relax
\mciteBstWouldAddEndPuncttrue
\mciteSetBstMidEndSepPunct{\mcitedefaultmidpunct}
{\mcitedefaultendpunct}{\mcitedefaultseppunct}\relax
\EndOfBibitem
\bibitem{Fan1}
X.~Fan, V.~K. Narayanan, R.~H. Lupton, M.~A. Strauss, G.~R. Knapp {\em et~al.}, \ifthenelse{\boolean{articletitles}}{\emph{A survey of z $>$ 5.8 quasars in the sloan digital sky survey. i. discovery of three new quasars and the spatial density of luminous quasars at z$\sim$ 6}, }{}The Astronomical Journal \textbf{122} (2001) 2833\relax
\mciteBstWouldAddEndPuncttrue
\mciteSetBstMidEndSepPunct{\mcitedefaultmidpunct}
{\mcitedefaultendpunct}{\mcitedefaultseppunct}\relax
\EndOfBibitem
\bibitem{Ferrarese_Msigma}
L.~Ferrarese and D.~Merritt, \ifthenelse{\boolean{articletitles}}{\emph{A fundamental relation between supermassive black holes and their host galaxies}, }{}The Astrophysical Journal \textbf{539} (2000) L9\relax
\mciteBstWouldAddEndPuncttrue
\mciteSetBstMidEndSepPunct{\mcitedefaultmidpunct}
{\mcitedefaultendpunct}{\mcitedefaultseppunct}\relax
\EndOfBibitem
\bibitem{inayoshiReview}
K.~Inayoshi, E.~Visbal, and Z.~Haiman, \ifthenelse{\boolean{articletitles}}{\emph{The assembly of the first massive black holes}, }{}Annual Review of Astronomy and Astrophysics \textbf{58} (2020) 27\relax
\mciteBstWouldAddEndPuncttrue
\mciteSetBstMidEndSepPunct{\mcitedefaultmidpunct}
{\mcitedefaultendpunct}{\mcitedefaultseppunct}\relax
\EndOfBibitem
\bibitem{volonteriReview}
M.~Volonteri, M.~Habouzit, and M.~Colpi, \ifthenelse{\boolean{articletitles}}{\emph{The origins of massive black holes}, }{}Nature Reviews Physics \textbf{3} (2021) 732\relax
\mciteBstWouldAddEndPuncttrue
\mciteSetBstMidEndSepPunct{\mcitedefaultmidpunct}
{\mcitedefaultendpunct}{\mcitedefaultseppunct}\relax
\EndOfBibitem
\bibitem{Turner1991quasars}
E.~L. Turner, \ifthenelse{\boolean{articletitles}}{\emph{Quasars and galaxy formation. i-the z greater than 4 objects}, }{}Astronomical Journal (ISSN 0004-6256), vol.\ 101, Jan.\ 1991, p.\ 5-17.\  \textbf{101} (1991) 5\relax
\mciteBstWouldAddEndPuncttrue
\mciteSetBstMidEndSepPunct{\mcitedefaultmidpunct}
{\mcitedefaultendpunct}{\mcitedefaultseppunct}\relax
\EndOfBibitem
\bibitem{LaceyCole}
C.~{Lacey} and S.~{Cole}, \ifthenelse{\boolean{articletitles}}{\emph{{Merger rates in hierarchical models of galaxy formation}}, }{}\href{https://doi.org/10.1093/mnras/262.3.627}{\mnras \textbf{262} (1993) 627}\relax
\mciteBstWouldAddEndPuncttrue
\mciteSetBstMidEndSepPunct{\mcitedefaultmidpunct}
{\mcitedefaultendpunct}{\mcitedefaultseppunct}\relax
\EndOfBibitem
\bibitem{fakhouri2010merger}
O.~Fakhouri, C.-P. Ma, and M.~Boylan-Kolchin, \ifthenelse{\boolean{articletitles}}{\emph{The merger rates and mass assembly histories of dark matter haloes in the two millennium simulations}, }{}Monthly Notices of the Royal Astronomical Society \textbf{406} (2010) 2267\relax
\mciteBstWouldAddEndPuncttrue
\mciteSetBstMidEndSepPunct{\mcitedefaultmidpunct}
{\mcitedefaultendpunct}{\mcitedefaultseppunct}\relax
\EndOfBibitem
\bibitem{Desjacques:2016bnm}
V.~Desjacques, D.~Jeong, and F.~Schmidt, \ifthenelse{\boolean{articletitles}}{\emph{{Large-Scale Galaxy Bias}}, }{}\href{https://doi.org/10.1016/j.physrep.2017.12.002}{Phys.\ Rept.\  \textbf{733} (2018) 1}, \href{http://arxiv.org/abs/1611.09787}{{\normalfont\ttfamily arXiv:1611.09787}}\relax
\mciteBstWouldAddEndPuncttrue
\mciteSetBstMidEndSepPunct{\mcitedefaultmidpunct}
{\mcitedefaultendpunct}{\mcitedefaultseppunct}\relax
\EndOfBibitem
\bibitem{Ferrarese:2002ct}
L.~Ferrarese, \ifthenelse{\boolean{articletitles}}{\emph{{Beyond the bulge: a fundamental relation between supermassive black holes and dark matter halos}}, }{}\href{https://doi.org/10.1086/342308}{Astrophys.\ J.\  \textbf{578} (2002) 90}, \href{http://arxiv.org/abs/astro-ph/0203469}{{\normalfont\ttfamily arXiv:astro-ph/0203469}}\relax
\mciteBstWouldAddEndPuncttrue
\mciteSetBstMidEndSepPunct{\mcitedefaultmidpunct}
{\mcitedefaultendpunct}{\mcitedefaultseppunct}\relax
\EndOfBibitem
\bibitem{Colin:2017juj}
J.~Colin, R.~Mohayaee, M.~Rameez, and S.~Sarkar, \ifthenelse{\boolean{articletitles}}{\emph{{High redshift radio galaxies and divergence from the CMB dipole}}, }{}\href{https://doi.org/10.1093/mnras/stx1631}{\mnras \textbf{471} (2017) 1045}, \href{http://arxiv.org/abs/1703.09376}{{\normalfont\ttfamily arXiv:1703.09376}}\relax
\mciteBstWouldAddEndPuncttrue
\mciteSetBstMidEndSepPunct{\mcitedefaultmidpunct}
{\mcitedefaultendpunct}{\mcitedefaultseppunct}\relax
\EndOfBibitem
\bibitem{Bengaly_2018}
C.~A.~P. Bengaly, R.~Maartens, and M.~G. Santos, \ifthenelse{\boolean{articletitles}}{\emph{Probing the cosmological principle in the counts of radio galaxies at different frequencies}, }{}\href{https://doi.org/10.1088/1475-7516/2018/04/031}{\jcap \textbf{2018} (2018) 031}\relax
\mciteBstWouldAddEndPuncttrue
\mciteSetBstMidEndSepPunct{\mcitedefaultmidpunct}
{\mcitedefaultendpunct}{\mcitedefaultseppunct}\relax
\EndOfBibitem
\bibitem{Secrest:2020has}
N.~J. Secrest, S.~v. Hausegger, M.~Rameez, R.~Mohayaee, S.~Sarkar {\em et~al.}, \ifthenelse{\boolean{articletitles}}{\emph{A test of the cosmological principle with quasars}, }{}\href{https://doi.org/10.3847/2041-8213/abdd40}{\apj \textbf{908} (2021) L51}\relax
\mciteBstWouldAddEndPuncttrue
\mciteSetBstMidEndSepPunct{\mcitedefaultmidpunct}
{\mcitedefaultendpunct}{\mcitedefaultseppunct}\relax
\EndOfBibitem
\bibitem{Siewert:2020krp}
T.~M. Siewert, M.~Schmidt-Rubart, and D.~J. Schwarz, \ifthenelse{\boolean{articletitles}}{\emph{Cosmic radio dipole: Estimators and frequency dependence}, }{}\href{https://doi.org/10.1051/0004-6361/202039840}{\aap \textbf{653} (2021) A9}\relax
\mciteBstWouldAddEndPuncttrue
\mciteSetBstMidEndSepPunct{\mcitedefaultmidpunct}
{\mcitedefaultendpunct}{\mcitedefaultseppunct}\relax
\EndOfBibitem
\bibitem{Secrest:2022uvx}
N.~J. Secrest, S.~von Hausegger, M.~Rameez, R.~Mohayaee, and S.~Sarkar, \ifthenelse{\boolean{articletitles}}{\emph{A challenge to the standard cosmological model}, }{}\href{https://doi.org/10.3847/2041-8213/ac88c0}{\apj { }Letters \textbf{937} (2022) L31}, \href{http://arxiv.org/abs/2206.05624}{{\normalfont\ttfamily arXiv:2206.05624}}\relax
\mciteBstWouldAddEndPuncttrue
\mciteSetBstMidEndSepPunct{\mcitedefaultmidpunct}
{\mcitedefaultendpunct}{\mcitedefaultseppunct}\relax
\EndOfBibitem
\bibitem{Grimm:2023tfl}
N.~Grimm, M.~Pijnenburg, S.~Mastrogiovanni, C.~Bonvin, S.~Foffa {\em et~al.}, \ifthenelse{\boolean{articletitles}}{\emph{{Combining chirp mass, luminosity distance, and sky localization from gravitational wave events to detect the cosmic dipole}}, }{}\href{https://doi.org/10.1093/mnras/stad3034}{Mon.\ Not.\ Roy.\ Astron.\ Soc.\  \textbf{526} (2023) 4673}, \href{http://arxiv.org/abs/2309.00336}{{\normalfont\ttfamily arXiv:2309.00336}}\relax
\mciteBstWouldAddEndPuncttrue
\mciteSetBstMidEndSepPunct{\mcitedefaultmidpunct}
{\mcitedefaultendpunct}{\mcitedefaultseppunct}\relax
\EndOfBibitem
\bibitem{Mastrogiovanni:2022nya}
S.~Mastrogiovanni, C.~Bonvin, G.~Cusin, and S.~Foffa, \ifthenelse{\boolean{articletitles}}{\emph{{Detection and estimation of the cosmic dipole with the einstein telescope and cosmic explorer}}, }{}\href{https://doi.org/10.1093/mnras/stad430}{Mon.\ Not.\ Roy.\ Astron.\ Soc.\  \textbf{521} (2023) 984}, \href{http://arxiv.org/abs/2209.11658}{{\normalfont\ttfamily arXiv:2209.11658}}\relax
\mciteBstWouldAddEndPuncttrue
\mciteSetBstMidEndSepPunct{\mcitedefaultmidpunct}
{\mcitedefaultendpunct}{\mcitedefaultseppunct}\relax
\EndOfBibitem
\bibitem{Cousins:2024bhk}
B.~Cousins, A.~Dhani, B.~S. Sathyaprakash, and N.~Yunes, \ifthenelse{\boolean{articletitles}}{\emph{{Finding cosmic anisotropy with networks of next-generation gravitational-wave detectors}}, }{}\href{http://arxiv.org/abs/2406.15550}{{\normalfont\ttfamily arXiv:2406.15550}}\relax
\mciteBstWouldAddEndPuncttrue
\mciteSetBstMidEndSepPunct{\mcitedefaultmidpunct}
{\mcitedefaultendpunct}{\mcitedefaultseppunct}\relax
\EndOfBibitem
\bibitem{limber1953analysis}
D.~N. Limber, \ifthenelse{\boolean{articletitles}}{\emph{The analysis of counts of the extragalactic nebulae in terms of a fluctuating density field.}, }{}Astrophysical Journal, vol.\ 117, p.\ 134 \textbf{117} (1953) 134\relax
\mciteBstWouldAddEndPuncttrue
\mciteSetBstMidEndSepPunct{\mcitedefaultmidpunct}
{\mcitedefaultendpunct}{\mcitedefaultseppunct}\relax
\EndOfBibitem
\bibitem{Bartelmann:1999yn}
M.~Bartelmann and P.~Schneider, \ifthenelse{\boolean{articletitles}}{\emph{{Weak gravitational lensing}}, }{}\href{https://doi.org/10.1016/S0370-1573(00)00082-X}{Phys.\ Rept.\  \textbf{340} (2001) 291}, \href{http://arxiv.org/abs/astro-ph/9912508}{{\normalfont\ttfamily arXiv:astro-ph/9912508}}\relax
\mciteBstWouldAddEndPuncttrue
\mciteSetBstMidEndSepPunct{\mcitedefaultmidpunct}
{\mcitedefaultendpunct}{\mcitedefaultseppunct}\relax
\EndOfBibitem
\bibitem{LoVerde:2008re}
M.~LoVerde and N.~Afshordi, \ifthenelse{\boolean{articletitles}}{\emph{{Extended Limber Approximation}}, }{}\href{https://doi.org/10.1103/PhysRevD.78.123506}{Phys.\ Rev.\ D \textbf{78} (2008) 123506}, \href{http://arxiv.org/abs/0809.5112}{{\normalfont\ttfamily arXiv:0809.5112}}\relax
\mciteBstWouldAddEndPuncttrue
\mciteSetBstMidEndSepPunct{\mcitedefaultmidpunct}
{\mcitedefaultendpunct}{\mcitedefaultseppunct}\relax
\EndOfBibitem
\bibitem{Martinelli:2021ahc}
M.~Martinelli, R.~Dalal, F.~Majidi, Y.~Akrami, S.~Camera {\em et~al.}, \ifthenelse{\boolean{articletitles}}{\emph{{Ultralarge-scale approximations and galaxy clustering: Debiasing constraints on cosmological parameters}}, }{}\href{https://doi.org/10.1093/mnras/stab3578}{Mon.\ Not.\ Roy.\ Astron.\ Soc.\  \textbf{510} (2022) 1964}, \href{http://arxiv.org/abs/2106.15604}{{\normalfont\ttfamily arXiv:2106.15604}}\relax
\mciteBstWouldAddEndPuncttrue
\mciteSetBstMidEndSepPunct{\mcitedefaultmidpunct}
{\mcitedefaultendpunct}{\mcitedefaultseppunct}\relax
\EndOfBibitem
\bibitem{Lesgourgues:2011re}
J.~Lesgourgues, \ifthenelse{\boolean{articletitles}}{\emph{{The Cosmic Linear Anisotropy Solving System (CLASS) I: Overview}}, }{}\href{http://arxiv.org/abs/1104.2932}{{\normalfont\ttfamily arXiv:1104.2932}}\relax
\mciteBstWouldAddEndPuncttrue
\mciteSetBstMidEndSepPunct{\mcitedefaultmidpunct}
{\mcitedefaultendpunct}{\mcitedefaultseppunct}\relax
\EndOfBibitem
\bibitem{Zu:2012am}
Y.~Zu, D.~H. Weinberg, E.~Rozo, E.~S. Sheldon, J.~L. Tinker {\em et~al.}, \ifthenelse{\boolean{articletitles}}{\emph{{Cosmological Constraints from the Large Scale Weak Lensing of SDSS MaxBCG Clusters}}, }{}\href{https://doi.org/10.1093/mnras/stu033}{Mon.\ Not.\ Roy.\ Astron.\ Soc.\  \textbf{439} (2014) 1628}, \href{http://arxiv.org/abs/1207.3794}{{\normalfont\ttfamily arXiv:1207.3794}}\relax
\mciteBstWouldAddEndPuncttrue
\mciteSetBstMidEndSepPunct{\mcitedefaultmidpunct}
{\mcitedefaultendpunct}{\mcitedefaultseppunct}\relax
\EndOfBibitem
\bibitem{Mead:2020qdk}
A.~J. Mead and L.~Verde, \ifthenelse{\boolean{articletitles}}{\emph{{Including beyond-linear halo bias in halo models}}, }{}\href{https://doi.org/10.1093/mnras/stab748}{Mon.\ Not.\ Roy.\ Astron.\ Soc.\  \textbf{503} (2021) 3095}, \href{http://arxiv.org/abs/2011.08858}{{\normalfont\ttfamily arXiv:2011.08858}}\relax
\mciteBstWouldAddEndPuncttrue
\mciteSetBstMidEndSepPunct{\mcitedefaultmidpunct}
{\mcitedefaultendpunct}{\mcitedefaultseppunct}\relax
\EndOfBibitem
\bibitem{Planck:2018vyg}
Planck, N.~Aghanim {\em et~al.}, \ifthenelse{\boolean{articletitles}}{\emph{{Planck 2018 results. VI. Cosmological parameters}}, }{}\href{https://doi.org/10.1051/0004-6361/201833910}{Astron.\ Astrophys.\  \textbf{641} (2020) A6}, \href{http://arxiv.org/abs/1807.06209}{{\normalfont\ttfamily arXiv:1807.06209}}, [Erratum: Astron.Astrophys. 652, C4 (2021)]\relax
\mciteBstWouldAddEndPuncttrue
\mciteSetBstMidEndSepPunct{\mcitedefaultmidpunct}
{\mcitedefaultendpunct}{\mcitedefaultseppunct}\relax
\EndOfBibitem
\bibitem{Babak:2024yhu}
S.~Babak, M.~Falxa, G.~Franciolini, and M.~Pieroni, \ifthenelse{\boolean{articletitles}}{\emph{{Forecasting the sensitivity of pulsar timing arrays to gravitational wave backgrounds}}, }{}\href{https://doi.org/10.1103/PhysRevD.110.063022}{Phys.\ Rev.\ D \textbf{110} (2024) 063022}, \href{http://arxiv.org/abs/2404.02864}{{\normalfont\ttfamily arXiv:2404.02864}}\relax
\mciteBstWouldAddEndPuncttrue
\mciteSetBstMidEndSepPunct{\mcitedefaultmidpunct}
{\mcitedefaultendpunct}{\mcitedefaultseppunct}\relax
\EndOfBibitem
\bibitem{Depta:2024ykq}
P.~F. Depta, V.~Domcke, G.~Franciolini, and M.~Pieroni, \ifthenelse{\boolean{articletitles}}{\emph{{Pulsar timing array sensitivity to anisotropies in the gravitational wave background}}, }{}\href{https://doi.org/10.1103/PhysRevD.111.083039}{Phys.\ Rev.\ D \textbf{111} (2025) 083039}, \href{http://arxiv.org/abs/2407.14460}{{\normalfont\ttfamily arXiv:2407.14460}}\relax
\mciteBstWouldAddEndPuncttrue
\mciteSetBstMidEndSepPunct{\mcitedefaultmidpunct}
{\mcitedefaultendpunct}{\mcitedefaultseppunct}\relax
\EndOfBibitem
\bibitem{Semenzato:2024mtn}
F.~Semenzato, J.~A. Casey-Clyde, C.~M.~F. Mingarelli, A.~Raccanelli, N.~Bellomo {\em et~al.}, \ifthenelse{\boolean{articletitles}}{\emph{{Cross-Correlating the Universe: The Gravitational Wave Background and Large-Scale Structure}}, }{}\href{http://arxiv.org/abs/2411.00532}{{\normalfont\ttfamily arXiv:2411.00532}}\relax
\mciteBstWouldAddEndPuncttrue
\mciteSetBstMidEndSepPunct{\mcitedefaultmidpunct}
{\mcitedefaultendpunct}{\mcitedefaultseppunct}\relax
\EndOfBibitem
\end{mcitethebibliography}

\appendix

\section{Pulsar-averaged galaxy clustering variance} \label{App:Shot_pulsar_variance}

In Section~\ref{AnalyticResults}, we have presented results on the variance of the HD correlation when considering a single pulsar pair at fixed positions $\bn_a$ and $\bn_b$. However, one can as well first average over all possible locations $\bn_a$ and $\bn_b$ separated by a fixed angle $\gamma$, see Ref.~\cite{Allen:2022dzg}, corresponding to the limit of a survey with infinitely many observed pulsars. Thereby, one obtains a different observable, with expectation value as well given by the standard HD curve but having a reduced variance. More precisely, the HD correlation $\rho_{ab}$ depending on the pulsars $a$ and $b$ is replaced by the pulsar averaged correlation $\rho(\gamma)$,\footnote{A possible source of confusion could arise from the fact that while $\rho(\gamma)$ depends only on the pulsar pair separation $\gamma$, this is true for the quantity $\rho_{ab}(\gamma)$ as well. Indeed, if we consider only one pulsar pair $a$ and $b$, we can choose a coordinate system such that the pulsar $a$ is aligned with the $z$-direction, while $b$ lies in the $x-z$ plane. However, the quantity $\rho(\gamma)$ represents an average over all possible pulsar pairs. As soon as more than one pair is involved, a unique choice of coordinate system does not allow to align one pulsar in each pair with the $z$-direction. Therefore, even though both $\rho(\gamma)$ and $\rho_{ab}(\gamma)$ only depend on $\gamma$, these two are distinct quantities.}
\begin{equation}
\rho(\gamma)=\angbr{\rho_{ab}}_p=\angbr{\overline{Z_aZ_b}}_p\,,
\end{equation}
where $\langle\dots\rangle_p$ denotes the integral over all sky directions $\bn_a$ and $\bn_b$ at a fixed angular separation. In Ref.~\cite{Allen:2022dzg}, the variance of $\rho(\gamma)$ is calculated in the absence of inhomogeneities.\footnote{In Ref.~\cite{Allen:2022dzg} the standard variance of $\rho(\gamma)$ (denoted by $\Gamma(\gamma)$ therein), i.e.~without taking the presence of inhomogeneities into account, is referred to as ``cosmic variance'', and the difference between the variance of $\rho_{ab}(\gamma)$ and $\rho(\gamma)$ is referred to as ``pulsar variance''. Here, to avoid ambiguity with large-scale structure terminology, we refer to the variance of $\rho(\gamma)$ as the pulsar-averaged variance, occurring after averaging over infinitely many pulsar pairs.} Here, we calculate the additional galaxy clustering contribution (see also Refs.~\cite{Allen:2024bnk, Agarwal:2024hlj} for similar recent work).

Following Ref.~\cite{Allen:2022dzg}, we omit the exponential term in the round brackets of Eq.~\eqref{Eq:Redshift}, since this term will be rapidly oscillating and average to zero for pulsar distances much larger than $1/f$. Then, $\rho(\gamma)$ reads: 
\begin{align}
&\rho(\gamma)=\angbr{\overline{Z_aZ_b}}_p \nonumber \\
    &=\sum_{A,A'}\int\mathrm df\int\mathrm df'\int\mathrm d\bn\int\mathrm d\bn' \angbr{F^A_a(\bn)F^{A'}_b(\bn')}_p h_A^\ast(f,\bn)h_{A'}(f',\bn')\mbox{sinc}\rbr{\pi(f-f')T}\,. \label{Eq:rhogamma}
\end{align}
For brevity of notation, one can define
\begin{equation}
    \mu_{AA'}\left(\gamma,\beta(\bn,\bn')\right) \equiv \angbr{F^A_a(\bn)F^{A'}_b(\bn')}_p\,,
\end{equation}
where $\beta(\bn,\bn')=\cos^{-1}(\bn\cdot\bn')$ denotes the separation angle between the two GW sources at $\bn$ and $\bn'$. This quantity reflects the fact that the antenna pattern functions depend on the exact geometrical constellation, i.e.~the positions of both pulsars and the two GW source directions. After varying the pulsar pair, at fixed angular separation $\gamma$, over the sky, this dependence is reduced to a dependence on the two separation angles $\gamma$ and $\beta$ only.

Starting from Eq.~\eqref{Eq:rhogamma}, we can repeat the procedure in Section~III of Ref.~\cite{Grimm:2024lfj}, replacing any factors $R_a^A(f,\bn)R^{A'}_b(f',\bn')$ by $\mu_{AA'}(\gamma, \beta(\bn,\bn'))$. For the standard terms, i.e.~without the impact of inhomogeneities, this leads to
\begin{align}
&\angbr{\rho^2}^\st=\frac 14 \iint \mathrm df\,\mathrm df'\bar S_h(f)\bar S_h(f')\iint\frac{\mathrm d\bn}{4\pi}\frac{\mathrm d\bn'}{4\pi}\,\sum_{A,A'}\mu_{AA}(\gamma,0)\mu_{A'A'}(\gamma,0) \nonumber \\
&+\frac 12 \iint \mathrm df\,\mathrm df'\bar S_h(f)\bar S_h(f')\,\mathrm{sinc}^2\rbr{\pi(f-f')T}\iint\frac{\mathrm d\bn}{4\pi}\frac{\mathrm d\bn'}{4\pi}\,\sum_{A,A'}\mu_{AA'}(\gamma,\beta(\bn,\bn'))\mu_{AA'}(\gamma,\beta(\bn,\bn'))\,, \label{Eq:irr_standard}
\end{align}
corresponding to Eq.~(C44) in Ref.~\cite{Allen:2022dzg}. For the integral over angles in the first line, we note that 
\begin{equation}
    \sum_A \mu_{AA}(\gamma,0)=\sum_A \angbr{F^A_a(\bn)F^{A}_b(\bn)}_p = \mu_{\rm HD}(\gamma)
\end{equation}
is the usual HD curve (corresponding to having one GW point source and averaging over pulsar pair positions, see Section~II in Ref.~\cite{Allen:2022dzg}). Therefore, the first line in Eq.~\eqref{Eq:irr_standard} vanishes when subtracting the square of the mean, and the second line constitutes the \emph{pulsar-averaged standard variance} (an analytical solution for the integral over angles is given in Eq.~(G11) of Ref.~\cite{Allen:2022dzg}).

Let us now apply the same considerations to the clustering terms. We obtain:
\begin{align}
&\angbr{\rho^2}^\clust{\cond}=\iint \mathrm dz\,\mathrm dz' G(z)G(z')\iint\frac{\mathrm d\bn}{4\pi}\frac{\mathrm d\bn'}{4\pi}\,\delta_g(\bn,z)\delta_g(\bn',z')\sum_{A,A'}\mu_{AA}(\gamma,0)\mu_{A'A'}(\gamma,0) \nonumber \\
&+2\iint \mathrm dz\,\mathrm dz'\,\Gamma(z,z')\iint\frac{\mathrm d\bn}{4\pi}\frac{\mathrm d\bn'}{4\pi}\,\delta_g(\bn,z)\delta_g(\bn',z')\sum_{A,A'}\mu_{AA'}(\gamma,\beta(\bn,\bn'))\mu_{AA'}(\gamma,\beta(\bn,\bn')) \,, \label{Eq:clust_pulsar_averaged1}
\end{align}
which we can further write as
\begin{align}
\angbr{\rho^2}^\clust{\cond}=&\mu^2_{\rm HD}(\gamma) \iint \mathrm dz\,\mathrm dz' G(z)G(z')\iint\frac{\mathrm d\bn}{4\pi}\frac{\mathrm d\bn'}{4\pi}\,\delta_g(\bn,z)\delta_g(\bn',z')\nonumber \\
&+\iint \mathrm dz\,\mathrm dz'\,\Gamma(z,z')\iint\frac{\mathrm d\bn}{4\pi}\frac{\mathrm d\bn'}{4\pi}\,\delta_g(\bn,z)\delta_g(\bn',z'){\big[\mu^2(\gamma,\beta(\bn,\bn'))+\mu^2(\pi-\gamma,\pi-\beta(\bn,\bn'))\big]}\,. \label{Eq:clust_pulsar_averaged}
\end{align}
Here $\mu(\gamma, \beta)$ is defined as
\begin{equation}
    \mu(\gamma,\beta) \equiv \mu_{++}(\gamma,\beta)+\mu_{\times\times}(\gamma,\beta)\,,
\end{equation}
and we have applied the parity transformation properties $\mu_{++}(\gamma,\beta)=\mu_{++}(\pi-\gamma,\pi-\beta)$ and $\mu_{\times\times}(\gamma,\beta)=-\mu_{\times\times}(\pi-\gamma,\pi-\beta)$. 
Taking the full ensemble average and performing a change of variables, this leads to
\begin{align}
\angbr{\rho^2}^\clust=&\frac 12\mu^2_{\rm HD}(\gamma) \iint \mathrm dz\,\mathrm dz' G(z)G(z')\int_0^\pi\mathrm d\beta\,\sin\beta\,\xi_g(\cos\beta,z,z') \nonumber \\
&+\frac{1}{2} \iint \mathrm dz\,\mathrm dz'\,\Gamma(z,z')\int_0^\pi\mathrm d\beta\,\sin\beta\,\xi_g(\cos\beta,z,z')\big[\mu^2\rbr{\gamma,\beta}+\mu^2(\pi-\gamma,\pi-\beta)\big]\,.
\label{eq:irredclustvariance}
\end{align}
We note that the first term, proportional to the squared HD curve itself, is only vanishing if
\begin{equation}
\int_0^\beta\mathrm d\beta\,\sin\beta\,\xi_g(\cos\beta,z,z')\neq 0\,,
\end{equation}
or equivalently (considering the first line in Eq.~\eqref{Eq:clust_pulsar_averaged}), if $\int\mathrm d\bn\,\delta_g(\bn,z)\neq 0$. This term would correspond to a monopole fluctuation in the galaxy overdensity, and we omit it in our calculation. Cosmological fluctuations at very large angular scales are predicted to be small in a $\Lambda$CDM universe. Moreover, a monopole fluctuation in the number of galaxies and thus the number of GW sources, increases the sky-averaged spectral density $\bar S_h$ and can be reabsorbed in the GW strain.\footnote{This is analogous to the usual procedure in galaxy surveys, where galaxy overdensity is defined with respect to the average number density observed on the sky. This procedure is absorbing any monopole fluctuation. Indeed, such monopole would be completely degenerate with the number density of galaxies that would form in an isotropic universe.} 
In other word, it increases or decreases the overall amplitude of the HD correlation at any pulsar pair separation, but does not alter its shape, which is reflected in the proportionality to $\mu_{\rm HD}(\gamma)^2$. We therefore focus on the second line, which, using the Limber approximation as before, can be simplified as
\begin{align}
&\angbr{\rho^2}^\clust= \frac{1}{2} \iint \mathrm dz\,\Gamma(z,z)\frac{H(z)}{c}\int_0^\pi\mathrm d\beta\,\sin\beta\big[\mu^2\rbr{\gamma,\beta}+\mu^2(\pi-\gamma,\pi-\beta)\big]\int\frac{k\,\mathrm dk}{2\pi}P_g(k,z)J_0(kr_z\beta)\,.
\label{eq:irredclustvariance_Limber}
\end{align}
This integral can be solved numerically, using the analytical expression for $\mu^2(\gamma,\beta)$ given in Eq.~(G5) of Ref.~\cite{Allen:2022dzg}.

 \section{Use of the astrophysical catalogs} \label{app:numerics}
We extract the GW luminosity in a representative comoving volume element of the universe following the merger tree philosophy of Refs.~\cat, for the three different scenarios outlined in Section~\ref{sec:catalogs}.

The merger tree algorithms of Refs.~\cat\ provide us lists of synthetic mergers events, with each event $i$ being associated to a merger redshift $z_m^i$ and a chirp mass $\M^i$. Each event is also attributed a weight $W^i$ (with dimension of inverse volume) which gives the catalog the interpretation of a box of comoving volume $V_c^\text{box} = 1\,\text{Mpc}^3$ whose content evolves over $z$. The weighting of events by $W^i$ follows from a Press-Schechter prescription, we refer to Refs.~\cat\ for more details.

Assuming that the catalog binaries are emitting GW in the PTA frequency band following a quasi-circular relative orbital motion and a standard cosmology, the pair $(z_m^i, \M^i)$ fully determines $f_s^i(z)$ and $\dd E_s^i(z)/\dd t_s$, that is, the whole history (over redshift $z$) of the frequency and power emission of source $i$, via the chirp equation \cite{Maggiore:1900zz}:
\be
\frac{\dd f^i_s}{ \dd t_s} =  \frac{96}{5} \pi^{8/3} \left(\frac{G\M^i}{c^3}\right)^{5/3} (f^i_s)^{11/3}\,.\label{eq:chirp}
\ee
We associate the local time of the source at redshift $z$ to be the cosmic time $t_s(z)$ of the Universe.

The Press-Schechter weights ensure that the box comoving volume at redshift $z$ is a fair representation of the Universe at that redshift. The sky-averaged GW luminosity density of the Universe is therefore assumed to be equal to the one of the box, which is given by the cumulative power of all catalog sources $i$ at redshift $z$, 
\begin{equation}
 \frac{\dd  \bar{L}_\text{GW}}{\dd V_c}(z) = \frac{L^\text{box}_\text{GW}(z)}{V_c^\text{box}} =	\sum_i W_i \frac{\dd E_s^i}{\dd t_s}(z)\,,
\end{equation}
containing the contribution of all sources regardless of their frequencies. This function may in practice not be very smooth, since, at each redshift $z$, it consists of a sum over the contributions of a discrete and possibly small amount of sources. We therefore smooth it by taking a (sliding) average at each redshift $z$. In the following, $\Delta t_s$ is the source time interval over which the smoothing average is taken:
\begin{align}
	\sum_i W_i \frac{\dd E_s^i}{\dd t_s} \simeq& \frac{1}{\Delta t_s}\int_{\Delta t_s} \dd t_s \sum_i W_i \frac{\dd E_s^i}{\dd t_s} = \frac{1}{\Delta t_s} \sum_i W_i \int_{\Delta f^i_s} \dd f_s^i \frac{\dd E_s^i}{\dd f_s^i} \\ =& \frac{\pi^{2/3}}{3 G} \frac{1}{\Delta t_s} \sum_i W_i (G\M^i)^{5/3} \int_{\Delta f^i_s} \dd f_s^i (f_s^i)^{-1/3} = \frac{\pi^{2/3}}{2 G} \frac{1}{\Delta t_s} \sum_i W_i (G\M^i)^{5/3} \left[(f_s^i)^{2/3}\right]^{\text{max($\Delta f^i_s$)}}_{\text{min($\Delta f^i_s$)}}\,, \label{eq:catalogL}
\end{align}
where, for each $i$, we changed the integration variable from the cosmic time at the source $t_s$ to $f_s^i$. Moreover, $\Delta f^i_s$ is the source frequency interval of integration corresponding to the so far generic smoothing time interval $\Delta t_s$. 
In practice, we use a smoothing cosmic time interval at the source such that it corresponds to a redshift interval $\Delta z = 0.1$, that is, $\Delta t_s(z) = |t_s(z+0.05)-t_s(z-0.05)|$.
We then refine the analysis with a logarithmic $\Delta z$ smoothing at $z < 0.05$ to capture, with a finer resolution, the most recent catalog events, since the luminosity must vanish at $z=0$ (no source is present directly at the observer).

The above luminosity contains the contributions of all frequencies.
The analysis is then divided into the 14 different observed frequency bands of the NANOGrav collaboration \cite{NANOGrav:2023gor}, that is, $f_j = j/T, \ j = 1,\dots, 14$, with $\Delta f = 1/T$ and $T$ the observation time.
For each catalog event $i$, we numerically solve the chirp equation \eqref{eq:chirp} to define
\be
\Delta z ^{i,j} = \left\{z : f_s^i(z)/(1+z) \in [ f_j-\Delta f/2, f_j+\Delta f/2]\right\}\,,
\ee
as the redshift interval over which source $i$ emits at a frequency that is observed in band $j$.

We then associate a source frequency interval $\Delta f_s ^{i,j}$ to $\Delta z ^{i,j}$ in the same way as we have previously associated a frequency interval $\Delta f_s^i$ to a generic source time interval $\Delta t_s$ or redshift interval $\Delta z$. If we then substitute
\be
\left[(f_s^i)^{2/3}\right]^{\text{max($\Delta f^i_s$)}}_{\text{min($\Delta f^i_s$)}}
\ee
by 
\be
 \sum_{j=1}^{14} \left[(f_s^i)^{2/3}\right]^{\text{max($\Delta f^i_s\cap\Delta f_s ^{i,j}$)}}_{\text{min($\Delta f^i_s\cap\Delta f_s ^{i,j}$)}} \,, \label{sumj}
\ee
in Eq.~\eqref{eq:catalogL}, we restrict the luminosity to the frequency contributions in the 14 observed PTA bands. The restriction to a single $j$-term in the sum \eqref{sumj} defines the luminosity in the observed band $j$.
\end{document}